\documentclass{aa}  
\usepackage{amsmath}
\usepackage{graphicx}
\usepackage{graphics}
\usepackage{subfigure}
\usepackage{txfonts}
\usepackage{natbib}
\bibpunct{(}{)}{;}{a}{}{,}

\def\teff{\ifmmode T_{\rm eff} \else $T_{\mathrm{eff}}$\fi}

\def\ltsima{$\buildrel<\over\sim$}
\def\lsim{\lower.5ex\hbox{\ltsima}~}

\begin{document}


   \title{On the detectability of Ly$\alpha$\ emission in star-forming galaxies \thanks{This work was supported by the Centre National d'Etudes Spatiales (CNES). It is based on observations made with the NASA/ESA Hubble Space Telescope, obtained at the Space Telescope Science Institute, which is operated by the Association of Universities for Research in Astronomy, Incorporated under NASA contract NAS 5-26555. These observations are associated with programs \#GO 9470 and \#GO 10575.} \thanks{Based on observations made with ESO Telescopes at the La Silla Observatories under programme IDs 073.B-0785 and 70.B-0639} \thanks{A part of data presented here have been taken using ALFOSC, which is owned by the Instituto de Astrofisica de Andalucia (IAA) and operated at the Nordic Optical Telescope under agreement between IAA and the NBIfAFG of the Astronomical Observatory of Copenhagen}  }

   \subtitle{The role of dust}

   \author{Hakim Atek\inst{1}$^\dagger$, Daniel Kunth\inst{1},  Matthew Hayes\inst{2}$^\ddagger$ , G\"oran \"Ostlin\inst{2} 
    \and  J. Miguel Mas-Hesse\inst{3}
 }

   \offprints{H. Atek}

\institute{Institut d'Astrophysique de Paris (IAP), 98bis boulevard Arago, 75014 Paris, France\\ 
         $^\dagger$\email{atek@iap.fr}
          \and 
             Stockholm Observatory, AlbaNova University Centre, 106 91 Stockholm, Sweden\\
         $^\ddagger$ present address: Observatoire de Gen\`eve, 51, Ch. des Maillettes, CH-1290, Sauverny, Switzerland   
         \and 
             Centro de Astrobiolog\'{\i}a (CSIC--INTA), 28850 Torrej\'on de Ardoz, Spain
             }

 \authorrunning{H. Atek et al.}
 \titlerunning{Ly$\alpha$\ emission in star-forming galaxies}

 \date{Received date; accepted date}

   \abstract
   {Lyman-alpha (Ly$\alpha$) is now widely used to investigate the galaxy formation and evolution in the high redshift universe.
. 
However, without a rigorous understanding of the processes which regulate the Ly$\alpha$\ escape fraction, physical interpretations of high-z observations remain questionable.
}
   {To examine six nearby star-forming galaxies to disentangle the role of the dust from other parameters such as gas kinematics, geometry and ISM morphology in the obscuration of Ly$\alpha$. Thereby we aim to understand the Ly$\alpha$\ escape physics and infer the implications for high-redshift studies.}
   {We use HST/ACS to produce continuum-subtracted Ly$\alpha$\ maps, and ground-based observations (ESO/NTT and NOT) to map the H$\alpha$\ emission and the extinction E(B-V) in the gas phase derived from the Balmer decrement H$\alpha$/H$\beta$.}
   { When large outflows are present, the Ly$\alpha$\ emission appears not to correlate with the dust content, confirming the role of the H{\sc i}\ kinematics in the escape of Ly$\alpha$\ photons. In the case of a dense, static H{\sc i}\ covering, we observe a damped absorption with a declining relationship between Ly$\alpha$\ and E(B-V).

We found that the Ly$\alpha$\ escape fraction does not exceed 10\% in all our galaxies and is mostly about 3\% or below. Finally, because of the radiative transfer complexity of the Ly$\alpha$\ line, star formation rate based on Ly$\alpha$\ luminosity is underestimated with respect to that derived from UV luminosity. Simple reddening correction does not reconcile SFR(Ly$\alpha$) with the total star formation rate.}
   { The dust is not necessarily  the main Ly$\alpha$\ escape regulatory factor. ISM kinematics and geometry may play a more significant role.
 
The failure of simple dust correction to recover the intrinsic
Ly$\alpha$/H$\alpha$\ ratio or the total star formation rate should prompt us to be more cautious when interpreting high-z observations and related properties, such as SFRs based on Ly$\alpha$\ alone. To this end we propose a more realistic calibration for SFR(Ly$\alpha$) which accounts for dust attenuation and resonant scattering effects via the Ly$\alpha$\ escape fraction.}


   \maketitle

\section{Introduction}
\paragraph{}
The Lyman-alpha emission (Ly$\alpha$) has become the most powerful tracer of star-formation in the high-redshift universe. It becomes the strongest emission line in the optical and near infrared (NIR) domain at redshift $z\geq2.1$, and is likely to remain a very competitive tool, even with the advent of extremely large telescopes (ELTs) and the James Webb Space Telescope (JWST). It is widely used as an efficient detection and redshift confirmation tool for distant galaxies, to derive star formation rates (SFRs), as well as to probe the ionisation state of the intergalactic medium (IGM) at the final stage of the reionisation epoch. In this way, the last decade has been the \textit{high-redshift era}, in which the development of new techniques and facilities have allowed the exploration of the properties of galaxy population, resulting in a major improvement in our understanding of the distant Universe.

The importance of the Ly$\alpha$\ emission line in the cosmological context was predicted very early by \cite{pp67}, who suggested that young high-z galaxies, undergoing their first star formation events, should be detectable thanks to their strong Ly$\alpha$\ emission. Unfortunately, the first attempts to detect such objects stand in contrast with those predictions. Indeed, initial surveys \citep[e.g.][]{pritchet89,djorgovski92,depropris93} failed to discover the predicted space density of Ly$\alpha$\ emitters. Unsuccessful campaigns and faint Ly$\alpha$\ fluxes were attributed to the dust attenuation coupled to the resonant scattering of the Ly$\alpha$\ line \citep[see][for a review]{pritchet94}. The first break-through came with the observations of \citet{cowie98} and \citet{hu98} which encouraged high-efficiency surveys for high-z LAEs detection. Most notably, two techniques are now used to detect high-z galaxies. The Lyman Break Technique \citep{steidel96} uses the absorption bluewards the Ly$\alpha$\ absorption edge to detect the so-called Lyman Break Galaxies (LBGs). Narrow-band imaging surveys that make use of the Ly$\alpha$\ recombination line produced by the reprocessed absorbed radiation to target Lyman-Alpha Emitters (LAEs). The success that has been attributed to these techniques has given Ly$\alpha$\ a key-role in the context of understanding the distant universe. Besides its use in the
identification of galaxies, it is used to put constraints on the cosmic reionisation \citep{malhotra04, kashikawa06,dijkstra07}, and to study the clustering properties and morphology of galaxies to the highest redshifts \citep{hamana04, ouchi05, murayama07}. Finally, it allows the estimation of the star formation rates at high redshift \citep{kudritzki00, fujita03, pirzkal07}.

 The Ly$\alpha$\ star formation rate is typically derived using the H$\alpha$\ calibration relation \citep{kennicutt98} and assuming a case B recombination theory for the Ly$\alpha$/H$\alpha$ line ratio. Nevertheless, SFRs inferred from the UV continuum are found to be inconsistent with SFR(Ly$\alpha$).  It appears that SFR(Ly$\alpha$) is typically lower than SFR(UV) by a factor of several \citep{ajiki03, taniguchi05, tapken07, gronwall07}. Correction for internal reddening may, in principle, reconcile these two indicators. Yet different extinctions experienced by the
continuum and the emission line may arise due to geometrical effects \citep{calzetti94, giavalisco96}, and radiative transfer effects of the Ly$\alpha$\ line make this issue far from being resolved. This demonstrates that caution should be taken when using such a calibration. This need for caution is further demonstrated in cosmological studies where only a fraction of UV-selected galaxies show Ly$\alpha$\ in emission \citep{shapley03}. Furthermore some high-z studies have unveiled very high rest frame Ly$\alpha$\ equivalent widths \citep[EWs,][]{rhoads03, finkelstein07} and it seems unlikely that such high EWs can result from the ionising output of a normal stellar population. In the case of inhomogeneous ISM, where the dust is distributed in neutral clouds, with an ionised inter-could medium, \citet{neufeld91}, and later \citet{hansen06} have shown that Ly$\alpha$\ photons could potentially escape easier than continuum radiation. In this scenario, intrinsic Ly$\alpha$\ EWs are expected to be enhanced, resulting in the large values observed in those studies.

Many efforts have been devoted to understanding the physical processes that govern the fraction of escaping Ly$\alpha$\ photons, regarding its potential cosmological importance. Early observations at low redshift \citep{meier81,deharveng86, terlevich93} have shown Ly$\alpha$\ to be much weaker than predictions or even absent from starburst galaxies. This weakening was first attributed to the dust attenuation, confirmed by the correlation observed between $EW_{\mathrm{Ly}\alpha}$\ and the metallicity \citep{charlot_fall93}. These interpretations notwithstanding, \citet{giavalisco96} reached the opposite
conclusion, finding no clear correlation between $EW_{\mathrm{Ly}\alpha}$\ or Ly$\alpha$/H$\beta$\ and E(B-V). Furthermore, dust correction failed to recover the intrinsic Ly$\alpha$/H$\beta$\ ratio predicted by case B recombination theory. Recent spectroscopic studies have outlined the Ly$\alpha$\ "observational puzzle". \citet{kunth94} and \citet{thuan97b}, with the Goddard High Resolution Spectrograph (GHRS) and the Space Telescope Imaging Spectrograph (STIS), have found damped Ly$\alpha$\ absorption in {\sc I}Zw 18 and \object{SBS 0335-052}, the most metal-deficient galaxies known at low-z. In the purely dust-regulated model, a prominent Ly$\alpha$\ emission feature would be expected. 
On the other hand \citet{lequeux95} have found strong Ly$\alpha$\ emission in a much more metal- and dust-rich starburst Haro 2. Further studies have shed light on the mechanisms by which Ly$\alpha$\ photons may escape their host. \citet{kunth98} observed Ly$\alpha$\ morphologies ranging from emission to absorption in 8 low-z starbursts. They found systematic blueshifts between Ly$\alpha$\ feature and Low Ionisation State (LIS) metal absorption features in the ISM when Ly$\alpha$\ is seen in emission, indicative of an outflowing neutral medium. P-cygni profiles, with a redshifted emission peak with respect to the systemic velocity, were also found in these spectra. Furthermore, \citet{mashesse03} made use of hydrodynamic models \citep{tenorio99} to propose a unified physical picture for the variety of observed Ly$\alpha$\ profiles as a function of the evolution of the burst and the viewing geometry. They found that the Ly$\alpha$\ emission visibility and shape are mostly driven by the kinematical configuration of the neutral gas. Likewise, over the past few years, theoretical works and numerical simulations have been developed with  the same purpose. \citet{ahn03} and \citet{verhamme06} have shown how the variety of Ly$\alpha$\ profiles are mostly driven by the expansion of a super-bubble of neutral gas and the properties of the ISM (H{\sc i}\ column density and dust content). \citet{hansen06} utilized the original idea of \citet{neufeld91} to investigate the effects of a multi-phase ISM. More cosmological-oriented simulations (Tasitsiomi 2006) have been carried out, although the effects of dust remain to be treated. The simulations are currently able to deal with arbitrary intrinsic emission characteristics, hydrogen density, velocity fields and dust distributions \citep{verhamme06}, and to consistently reproduce the Ly$\alpha$\ profiles observed in z $\sim$ 3 LBGs \citep{verhamme07, schaerer08}.

\begin{table*}[!htbp]
\begin{center}
\caption{Targets general properties. E(B-V)$_{\rm MW}$ is the Galactic extinction given by \citet{schlegel98} and M$_{\rm B}$ magnitudes are in AB system.
References: 
1: \citet{bergvall02}, 
2: \citet{papaderos06}, 
3: \citet{gonzalez98}, 
4: \citet{izotov01}  \label{targets}}
\begin{tabular}{lllllllll}
\hline\hline\\
Target & Other & RA(2000) & Dec(2000) &$E(B-V)_{\rm MW}$  & $z$ & $12+$ & $M_B$ & Ref \\ 
Name   &  Name &          &           &                   &     & $\log({\rm O/H})$ & &  \\  \\
\hline  \\
Haro 11         & ESO 350-38    & 00:36:52.5  &  -33:33:19  & 0.049 & 0.020598    & 7.9 & -20  & 1 \\
SBS 0335-052    & SBS 0335-052E & 03:37:44.0  &  -05:02:40  & 0.047 & 0.013486    & 7.3 & -17  & 2 \\
IRAS 08339+6517 & PGC 024283    & 08:38:23.2  &  +65:07:15  & 0.092 & 0.019113    & 8.7 & -21  & 3 \\
Tololo 65       & ESO 380-27    & 12:25:46.9  &  -36:14:01  & 0.074 & 0.009       & 7.6 & -15  & 4 \\
NGC 6090        & Mrk 496       & 16:11:40.7  &  +52:27:24  & 0.020 & 0.029304    & 8.8 & -21  & 3 \\
ESO 338-04      & Tol 1924-416  & 19:27:58.2  &  -41:34:32  & 0.087 & 0.009633    & 7.9 & -19  & 1 \\  \\

\hline
\end{tabular}
\end{center}
\end{table*}

\begin{table*}
\begin{minipage}[!h]{18cm}
\begin{center}
\begin{tabular}{lllllllll}
\hline\hline\\
Target          &\hspace{1.5cm}H$\alpha$\   &  & \hspace{1cm}H$\alpha$\ continuum &      &    \hspace{1.5cm}H$\beta$\  &  &\hspace{1cm}H$\beta$\ continuum &  \\ \\
                &Camera/Filter& Exp. & Camera/Filter & Exp. & Camera/Filter &Exp. & Camera/Filter       & Exp.  \\ 
ESO/NTT observations \\
\hline 
Haro 11         & EMMI-R/598  & 900   & EMMI-R/597  & 1200  & SuSI2/549 & 2866 & EMMI-R/770 & 1800    \\
SBS 0335-052    & EMMI-R/598  & 1800  & EMMI-R/596  & 4800  & SuSI2/548 & 900  & EMMI-R/771 & 4500    \\
ESO 338-04      & EMMI-R/597  & 1800  & SuSI2/778   & 3600  & SuSI2/719 & 2400 & EMMI-R/771 & 4500    \\ 
Tololo 65       & SuSI2/709  & 1200   & SuSI2/778   &  1200 & SuSI2/719 & 3600  & SuSI2/717 & 1800     \\ \\
NOT observations \\
\hline 
IRAS 08339+6517 & ALFOSC/70  & 2400   & ALFOSC/78   &  3000 & ALFOSC/113& 5000  &ALFOSC/17  & 3300     \\
NGC 6090        & ALFOSC/53  & 4200   & ALFOSC/78   &  2400 & ALFOSC/40 & 3600  &ALFOSC/17  & 2100      \\
\hline
\end{tabular}
\caption{Ground-based observations of our six targets. Northern targets have been observed with the Nordic Optical Telescope and southern ones with ESO New Technology Telescope. H$\alpha$, H$\beta$\ online and respective continuum observations are listed with the instrument and filter name and the exposure time (in seconds) in each band.}
\label{ground_obs}
\end{center}
\end{minipage}
\end{table*}

The complex nature of the Ly$\alpha$\ escape probability revealed by low-z spectroscopic studies, gave a rise to additional issues. The resonant scattering phenomenon of Ly$\alpha$\ line may cause Ly$\alpha$\ photons to travel and to be emitted far from their production sites hence to be spatially uncorrelated with non-resonant radiation (H$\alpha$\ or continuum photons). UV-targeted spectroscopic studies may therefore miss a significant fraction of the Ly$\alpha$\ emission if scattered away from the
aperture. Ionised holes in the ISM may also allow the escape of Ly$\alpha$\ photons in a spatially limited region, and transmission may vary greatly on small scales across the starburst region. These consideration are the motivation for our Ly$\alpha$ imaging survey with the Advanced Camera for Survey (ACS) on board the Hubble Space Telescope (HST). We have observed a hand-picked sample of six nearby star-forming galaxies, with the aim of exploring a large range of relevant parameters. Preliminary results have been presented in \citet{kunth03}, and more detailed studies on ESO 338-04
\citep{hayes05} and Haro 11 \citep{hayes07a}. Emission and absorption were found on very small scales in central regions of the starburst while absorption is observed in front of many of the brightest UV sources.

In this article, besides the HST observations, we make use of ground-based observations from the ESO New Technology Telescope (NTT) and the Nordic Optical Telescope (NOT) to map H$\alpha$\ and H$\beta$\ emission and derive the extinction from the Balmer decrement (H$\alpha$/H$\beta$). Investigating correlations between E(B-V) and Ly$\alpha$\ emission enables us to disentangle the role of the dust from other parameters in the Ly$\alpha$\ escape mechanism, and investigate the implications of proper dust correction on high-redshift studies. The paper is structured as follows: In Section \ref{data_sec} we describe the observations and the data reduction, in Section \ref{analysis_sec} we present the results, Section \ref{discussion_sec} is dedicated to the discussion of these results and finally in Section \ref{conclusion_sec} we present our conclusions. We assume throughout this paper a cosmology of $H_0$=72 km~s$^{-1}$~Mpc$^{-1}$, $\Omega_{M}$=0.3 and $\Omega_{\Lambda}$=0.7.

\section{OBSERVATIONS AND DATA REDUCTION}
\label{data_sec}

\paragraph{}
Our sample contains six local starburst galaxies that have been hand-picked in order to cover a range of relevant intrinsic parameters. Based upon UV-properties, the selection covers a range of dust content, luminosity and the variety of observed Ly$\alpha$\ profiles: firstly, the sample consists of four Ly$\alpha$\ emitters from \citet{calzetti92}, among which candidates
representing a range in line profiles are selected \citep{kunth98,
mashesse03}. Two known Ly$\alpha$\ absorbers are also included from
\citet{thuan97b}. General information and properties are given in Table
\ref{targets}

\subsection{HST observations}
\label{HST_section}
\paragraph{}
Two general observer programmes, with the Hubble Space Telescope (HST), have been devoted for the observations of the six targets: GO9470 that makes use of the Solar Blind Channel (SBC) of the Advanced Camera for Surveys (ACS) in order to observe Ly$\alpha$\ online (F122M) and continuum (F140LP), and GO10575 using the High Resolution Camera (HRC) and Wide Field Camera (WFC) to obtain H$\alpha$, near-UV and optical continuum broad band observations. Detailed description of these observations is presented in \citet{ostlin08}.  

All images are 'drizzled' using the MULTIDRIZZLE task in STSDAS package under NOAO/IRAF onto the same pixel scale (0.025\arcsec\ pix$^{-1}$) and the same orientation. Remaining shifts are corrected with GEOMAP and GEOTRAN, and cosmic rays removed using CREDIT task. The images for each target are then convolved to the same Point Spread Function (PSF) using DIGIPHOT/DAOPHOT, with the worse PSF image as reference. 
\paragraph{}
The production of continuum subtracted Ly$\alpha$\ images from F122M (online) and F140LP (offline) is not trivial and needs sophisticated techniques for many reasons. The effective wavelength of the continuum filter is rather far from the online filter ($\Delta\lambda/\lambda \simeq$ 0.22) and the UV continuum between the two filters deviates significantly from a power law ($f_{\lambda} \propto \lambda^{\beta}$) and is sensitive to age and E(B-V). Neglecting these considerations and describing the behavior of the UV continuum near Ly$\alpha$\ with a power law yielded results ranging from absorption to strong emission according to the values set  to $\beta$. Preliminary results, by \citet{kunth03}, pointed out these limitations and the need for an elaborated subtraction method. Subsequently, \citet{hayes05} and, in more details, \citet{hayes08} presented a reliable extrapolation from F140LP to F120M described by the Continuum Throughput Normalization factor (CTN). To estimate the CTN in each pixel, all the images from F122M to F814W are used to fit the \textit{Starburst99} spectral evolutionary models \citep{leitherer99, vasquez05}. Summarily, the method uses filters that sample the UV continuum slope and the 4000\AA\ break to fit burst age and stellar E(B-V) using ~$\chi^2$ minimisation. Then the nebular contribution and the stellar components are treated independently. Indeed, H$\alpha$\ image is used to constrain the nebular gas contribution to the total SED, while V and I images allow to estimate the stellar part. The gas spectrum is subtracted and age and mass are fitted in two stellar components.

 To generate Starburst99 models, we set the metallicity as derived from observations of each object: for \object{IRAS 08339+6517 } and \object{NGC 6090} $Z=0.02$, for \object{Haro 11} and \object{ESO 338-04} $Z=0.004$, and for SBS 335-052 and \object{Tololo 65} $Z=0.001$. Though \citet{hayes08} show that an error below 50\% on the metallicity does not significantly affect the continuum subtraction, the metallicity estimate remains the factor driving the accuracy of the subtraction method.

A standard \citet{salpeter55} IMF is used ($\alpha=2.35, dN=M^{-\alpha} dM$) in the range 0.1$M_\odot$ to 120$M_\odot$. Multi-stellar component fitting, used for continuum subtraction, is almost completely insensitive to the IMF slope and mass range. Finally, an instantaneous burst scenario is adopted as it is more appropriate for individual pixels but the choice of a constant star formation scenario would not affect global photometry of the galaxies or the contribution of the underlying stellar population.

\subsection{ground-based observations}
\label{ground_obs_sec}

\paragraph{}
The southern targets of our sample (cf. Table \ref{ground_obs}) have been observed using the New Technology Telescope (NTT) at La Silla (ESO) during the nights of 18, 19 and 20 September 2004 (apart from Tololo 65 which was observed in service mode on 28 January 2003). Narrow-band imaging was performed for all targets in H$\alpha$, H$\beta$\ and [O{\sc iii}]$\lambda 5007$\AA, and nearby continuum for each line. The first night of the run, 18th Sept, the seeing was consistently below 1.2\arcsec, yet the presence of thin cirrus prevents a direct calibration using standard stars. Observational conditions on the night of 19 Sept were excellent: photometric sky and sub-arcsec seeing. On the last night the seeing was worse, exceeding 2\arcsec, although the photometric quality was still good. Spectrophotometric standard stars Feige110, LDS749B, GD50 and GD108, selected from \citet{oke90} catalog, were observed at regular intervals during each night and in each filter. Both the ESO Multi-Mode Instrument (EMMI) \citep{dekker86} and Super Seeing Imager 2 (SuSI2) \citep{dodorico98} were used interchangeably. A binning of 2$\times$2 pixels was used, giving plate-scales of 0.332\arcsec\ pix$^{-1}$ and 0.161\arcsec\ pix$^{-1}$, and fields-of-view of 9.1 $\times$ 9.9\arcmin\ and 5.5 $\times$ 5.5\arcmin\, for EMMI-R and SuSI2 respectively. NTT observations are summarized in table \ref{ground_obs} together with ESO filter codes. The good seeing observations of the night 18 Sept are calibrated using secondary standard stars in the field from the photometric night 19 and/or 20 Sept.

The remaining northern targets (IRAS 08339+6517  and NGC 6090) were observed with the Nordic Optical Telescope (NOT) at La Palma during the nights of 14, 15 and 16 February 2004, using the Andalucia Faint Object Spectrograph and Camera (ALFOSC). Unfortunately, the observational conditions were not photometric during the observing run. For NGC 6090, we used observations taken on March 28th 2006 under photometric conditions. The calibration was performed using BD33 standard star. Our science images were then calibrated using field stars from 2006 observations in each filter. For IRAS 08339+6517 , in the same purpose, we made use of HST images. Firstly, calibrated H$\alpha$\ images are rescaled to the ALFOSC pixel-plate (0.19\arcsec\ pix$^{-1}$). Aperture photometry is then obtained and used for the calibration of the ground-based H$\alpha$\ observations. The H$\beta$\ continuum images are calibrated using rescaled F435W images and H$\beta$\ online images are then calibrated with respect to the continuum images using photometry of several stars in the field of view.  Global estimates for IRAS 08339+6517 are the least accurate of the galaxy sample due to the less-than-ideal calibration of H$\beta$. However, pixel-to-pixel measurements suffer from similar uncertainties, therefore pixel-to-pixel trends are as reliable as they are in any other target, irrespective whether the exact E(B-V) is perfect. 

\paragraph{}
Data were first reduced using the standard NOAO/IRAF procedures. Images were bias-subtracted, and flat-field corrected using well exposed sky and dome flats. Images were registered using field stars and the GEOMAP and GEOTRAN tasks, and images from the same bands were coadded with inverse variance weighting. Finally, all reduced frames for each target were smoothed to the PSF of the worst seeing image using GAUSS.

H$\alpha$\ images were corrected for [N{\sc ii}] contamination and both H$\alpha$\ and H$\beta$\ images corrected for underlying stellar absorption. [N{\sc ii}] contamination is estimated from N{\sc ii}\ $\lambda 6583\AA/\mathrm{H}\alpha$ found in literature (see Table \ref{nii}) and the ratio N{\sc ii}\ $\lambda 6584 \AA$/N{\sc ii}\ $\lambda 6548 \AA \simeq 3$ \citep{osterbrock89}.

Stellar absorption is estimated using the best-fitting \textit{Starburst99} spectrum given by the procedure described in Sec. \ref{HST_section}. Latest stellar atmospheres models are used in \textit{Starburst99} stellar libraries \citep{martins05}, including full line-blanketing, which allows the best estimate of equivalent width of the H$\alpha$\ and H$\beta$\ stellar absorption features.
 
Finally, using the Balmer decrement H$\alpha$/H$\beta$, assuming a case B intrinsic line ratio of 2.86 \citep{osterbrock89} at a temperature of 10,000 K, we can derive the E(B-V) in the gas phase using the parameterization of \citet{cardelli89} and the equation:
\begin{equation}
 E(B-V)_{H\alpha/H\beta} = \frac{2.5 \times log(2.86 / R_{obs})}{k(\lambda_{\alpha}) - k(\lambda_{\beta})}
\end{equation}
where $R_{obs} = f_{H\alpha}/f_{H\beta}$\ is the absolute observed flux ratio, and k($\lambda_{\alpha}$), k($\lambda_{\beta}$) are the extinction curves at H$\alpha$ and H$\beta$ wavelengths respectively. According to \citet{cardelli89}, k($\lambda_{\alpha}$) $\sim$ 2.63 and k($\lambda_{\beta}$) $\sim$ 3.71.

\begin{table}[!htbp] 
\centering
\begin{tabular}{l c l }
\hline \hline \\
Target           & $\frac{\mathrm{[NII]}~\lambda 6583\AA}{\mathrm{H}\alpha}$ & Reference  \\  \\
\hline \\
Haro 11          &   0.189                             & \citet{bergvall02}    \\
ESO 338-04       &   0.031                             & \citet{bergvall02}   \\
NGC 6090         &   0.411                             & \citet{moustakas06}   \\
IRAS 08339+6517  &   0.25                              & \citet{margon88}  \\
SBS 0335-052     &   0.003                             & \citet{izotov97}  \\
Tololo 65        &   0.005                             & \citet{izotov04} \\
\hline \\
\end{tabular}
\caption{ H$\alpha$\ correction for N{\sc ii} contamination with references where the ratio has been taken from.} 
\label{nii} 
\end{table}

\subsection{Uncertainties}
\label{error_sec}
\paragraph{}
The calibration of individual images in H$\alpha$\ and H$\beta$\ lines are subject to classical errors in the data reduction procedure. The origin of the uncertainty comes from the flux measurement of the calibration standard stars. This accounts for about 5\% in our observations. Errors of the order of few percent may be produced by other effects, such as residuals in the flat-field corrections, residuals in the sky background and continuum subtraction. We make the assumption that all these effects will lead to an error on the flux measurement of the order of 10\%. Moreover, we are producing extinction maps by dividing H$\alpha$\ by H$\beta$\ images, and, for some galaxies, these images are obtained with different instruments (i.e. different plate-scales and orientations), the errors on images alignment and registration may lead to quite significant uncertainties, since we aim, in our science analysis, to investigate the Ly$\alpha$\ and dust amount variations on a pixel scale. The procedure to estimate the impact of such misalignments was the following:

The \textit{RMS} of the alignment fit given by GEOMAP is about 0.2 pixel. Consequently, to estimate the misalignment (and only to that purpose), we have rebinned H$\alpha$\ and H$\beta$\ images to a new pixel size (1/5 of the original pixel size). Then we have created new images with artificial shifts by +1,0 or -1 pixel in x and y directions. This gives us 9-image data cube for both H$\alpha$ and H$\beta$. Using all combinations of H$\alpha$\ and H$\beta$\ shifted images we derived the ratio H$\alpha$/H$\beta$ and constructed an E(B-V) data cube. We eventually computed the standard deviation of the extinction, $\sigma_{E(B-V)}$, at each pixel [x,y].

Misalignment errors are also a concern for the Ly$\alpha$\ image with respect to H$\alpha$ and online images with respect to continuum ones. This could result in errors on the Ly$\alpha$/H$\alpha$\ ratio and the equivalent widths. The estimation of these quantities takes into account this uncertainty. Besides, we have accounted for the uncertainties of the Ly$\alpha$\ continuum subtraction procedure by assuming a statistical error of 10\% of the Ly$\alpha$\ flux. This uncertainty depends, in reality, on the signal-to-noise in each resolution element where the continuum is fitted, and the value adopted here corresponds to S/N=5\%. The final error bars are shown on the scatter plots in Sect. \ref{analysis_sec}. For the sake of readability we plot the mean errors instead of individual errors for each point. When the emission is dominated by the conglomerate points of the diffuse halo (which is the case for most of the plots), we show rather the mean computed between only the maximum and minimum values, in order to cover the full range of error variations.

\section{Analysis}
\label{analysis_sec}

\begin{figure*}[!htbp]
 \centering
 \hspace{0.65cm}
 \includegraphics[width=5cm,height=0.5cm]{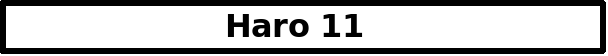}
 \hspace{0.1cm}
 \includegraphics[width=5cm,height=0.5cm]{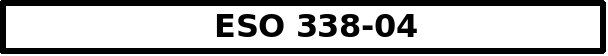}
 \hspace{0.1cm}
 \includegraphics[width=5cm,height=0.5cm]{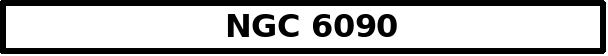}

 \vspace{0.1cm}

 \includegraphics[width=0.5cm,height=5cm]{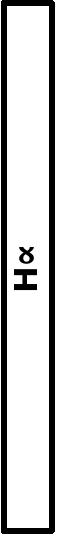}
 \hspace{0.1cm}
 \includegraphics[width=5cm,height=5cm]{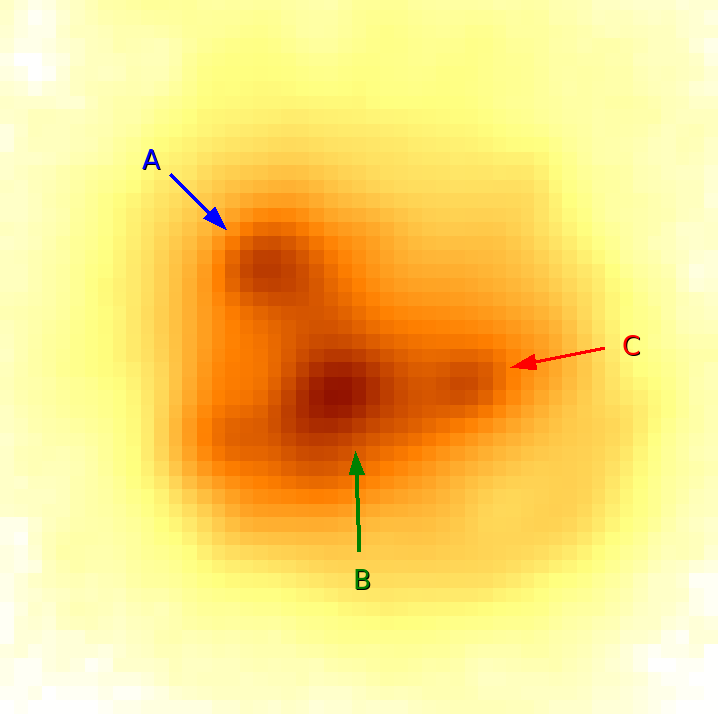}
 \hspace{0.1cm}
 \includegraphics[width=5cm,height=5cm]{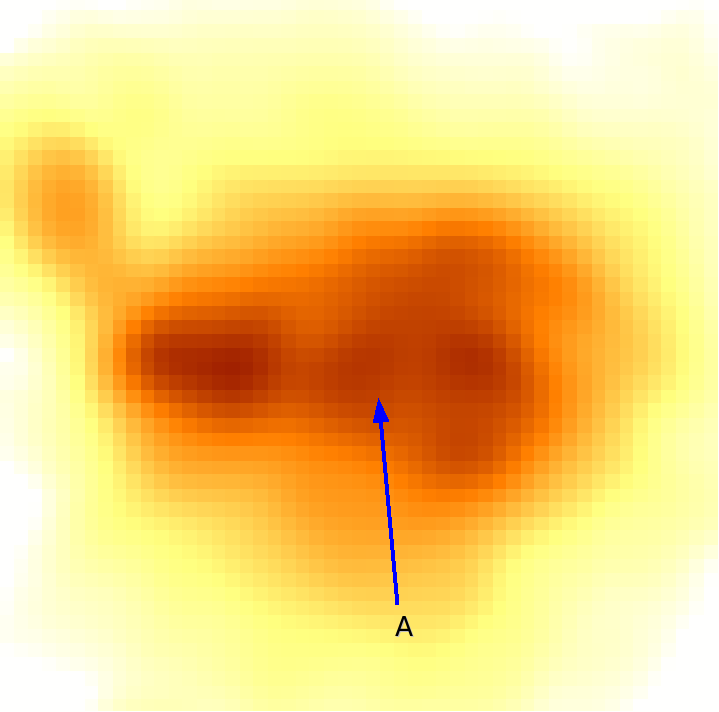}
 \hspace{0.1cm}
 \includegraphics[width=5cm,height=5cm]{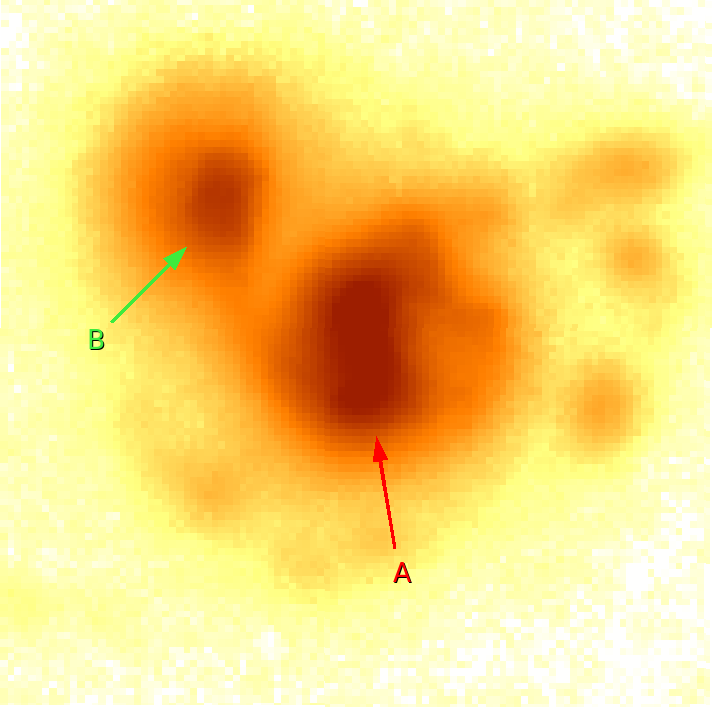}

 \vspace{0.3cm}

 \includegraphics[width=0.5cm,height=5cm]{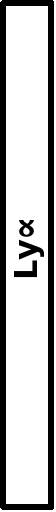}
 \hspace{0.1cm}
 \includegraphics[width=5cm,height=5cm]{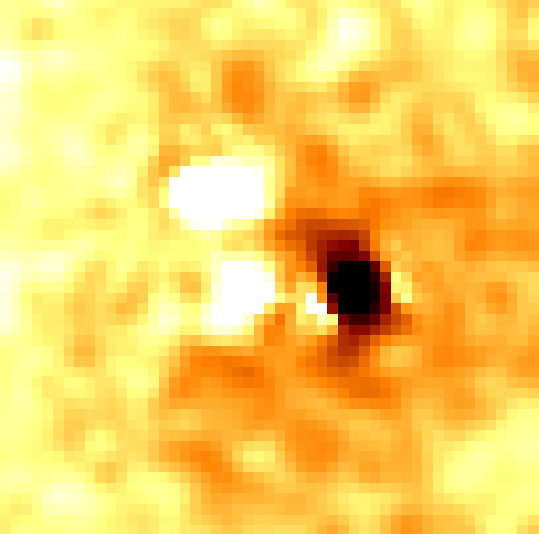}
 \hspace{0.1cm}
 \includegraphics[width=5cm,height=5cm]{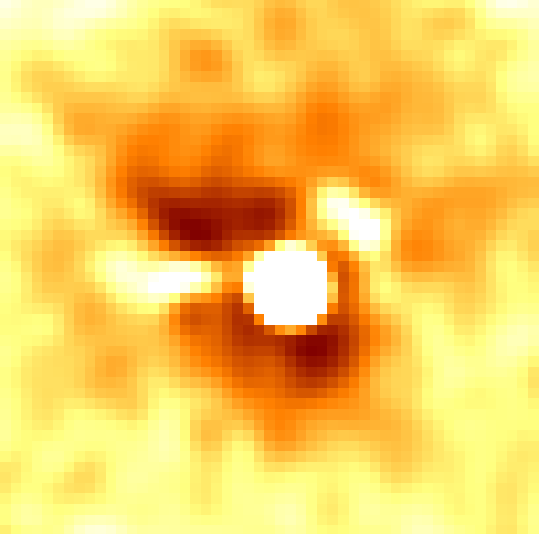}
 \hspace{0.1cm}
 \includegraphics[angle=90,angle=90,width=5cm,height=5cm]{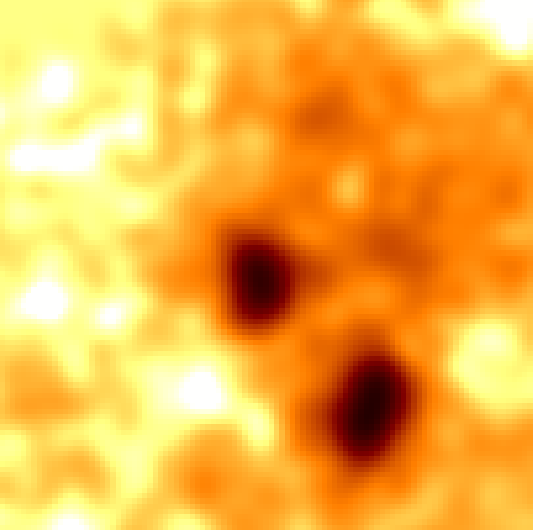}

 \vspace{0.3cm}
  
 \includegraphics[width=0.5cm,height=5cm]{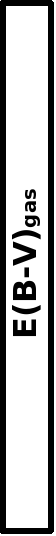}
 \hspace{0.1cm}
 \includegraphics[width=5cm,height=5cm]{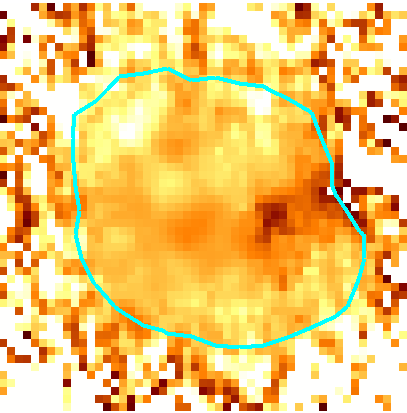}
 \hspace{0.1cm}
 \includegraphics[width=5cm,height=5cm]{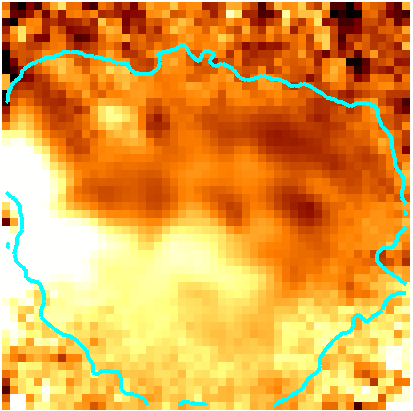}
 \hspace{0.1cm}
 \includegraphics[angle=90,angle=90,width=5cm,height=5cm]{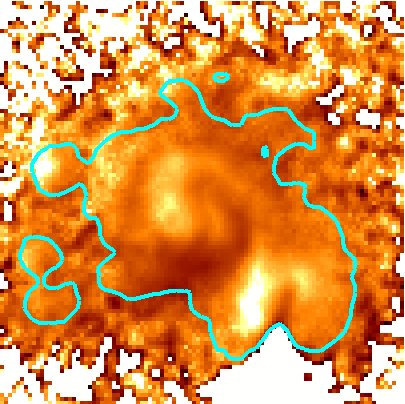}
  
 \vspace{0.3cm}

 \includegraphics[width=0.5cm,height=5cm]{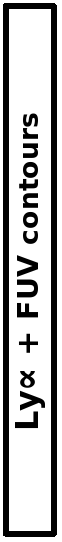}
 \hspace{0.1cm}
 \includegraphics[angle=90,angle=90,width=5cm,height=5cm]{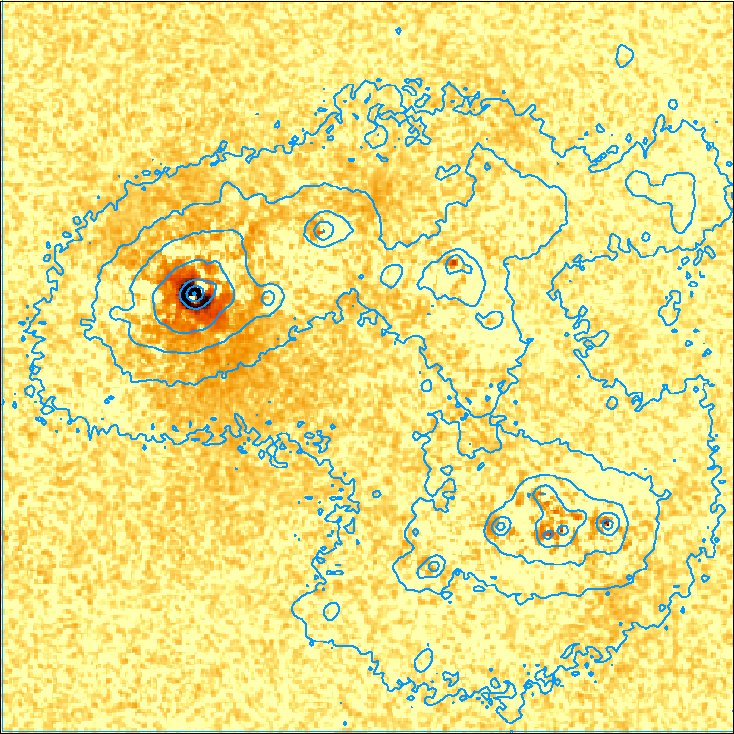}
 \hspace{0.1cm}
 \includegraphics[angle=90,angle=90,width=5cm,height=5cm]{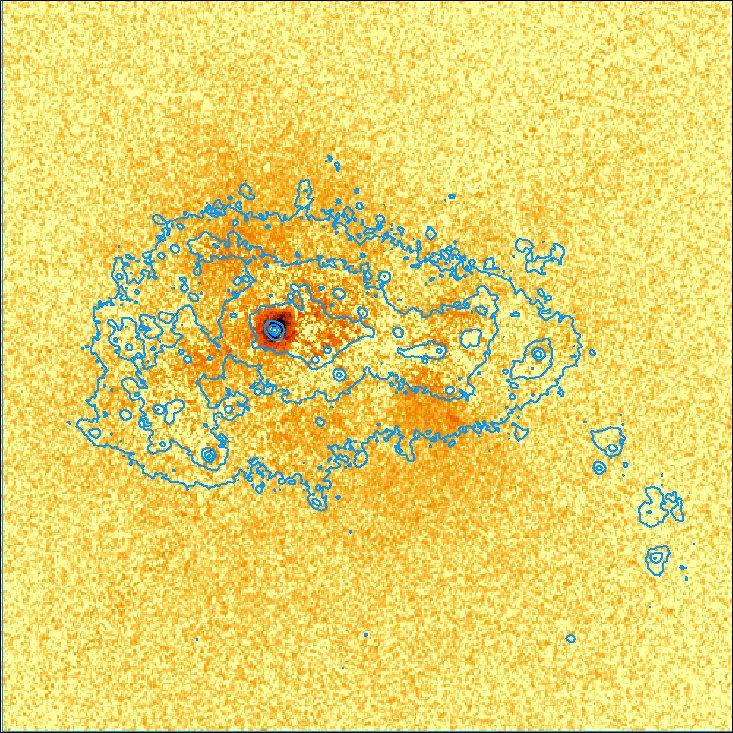}
 \hspace{0.1cm}
 \includegraphics[angle=90,angle=90,width=5cm,height=5cm]{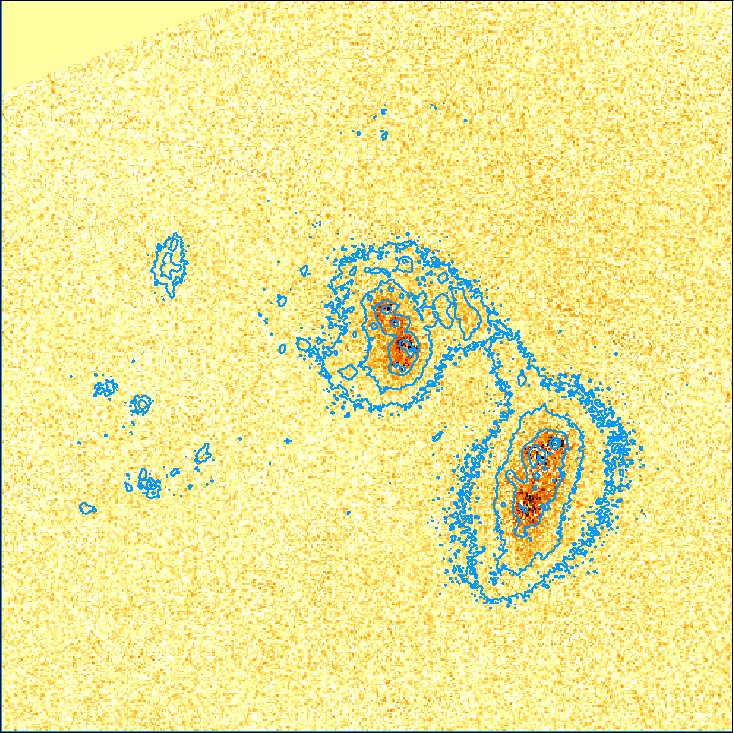}

  \vspace{0.5cm}

\caption{Galaxy imaging (Part 1): \textit{From top to bottom:} H$\alpha$\, Ly$\alpha$, E(B-V)$_{gas}$ map, and Ly$\alpha$\ as seen by HST overlayed with FUV contours. Inverted logarithmic scale is used, showing emission in black and absorption in white. The extinction map is overlayed with the mask generated following the description in Sect. \ref{discussion_sec}. Regions out of the contour are excluded from our study. The dustiest regions are in black. \textit{from left to right} with the FoV and spatial scale in parentheses: Ground-based images (top three rows): Haro 11 (17\arcsec, 0.4 kpc/\arcsec), ESO 338-04 (17\arcsec, 0.2 kpc/\arcsec), and NGC 6090 (19\arcsec, 0.57 kpc/\arcsec); HST images (last row): Haro 11 (13\arcsec), ESO 338-04 (15\arcsec), and NGC 6090 (21\arcsec). North is down and East to the right.} 
 \vspace{0.4cm}
 \label{images_1}
 \end{figure*}


\begin{figure*}[!htbp]
\centering
 \hspace{0.65cm}
 \includegraphics[width=5cm,height=0.5cm]{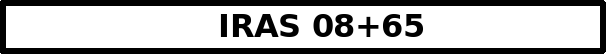}
 \hspace{0.1cm}
 \includegraphics[width=5cm,height=0.5cm]{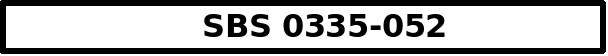}
 \hspace{0.1cm}
 \includegraphics[width=5cm,height=0.5cm]{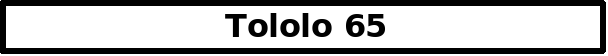}

  \vspace{0.1cm}

 \includegraphics[width=0.5cm,height=5cm]{images_2/ha_title.png}
 \hspace{0.1cm}
 \includegraphics[width=5cm,height=5cm]{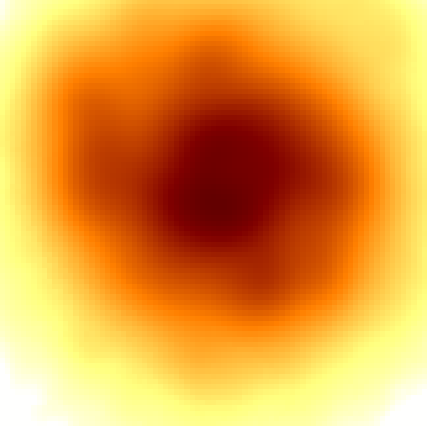}
 \hspace{0.1cm}
 \includegraphics[width=5cm,height=5cm]{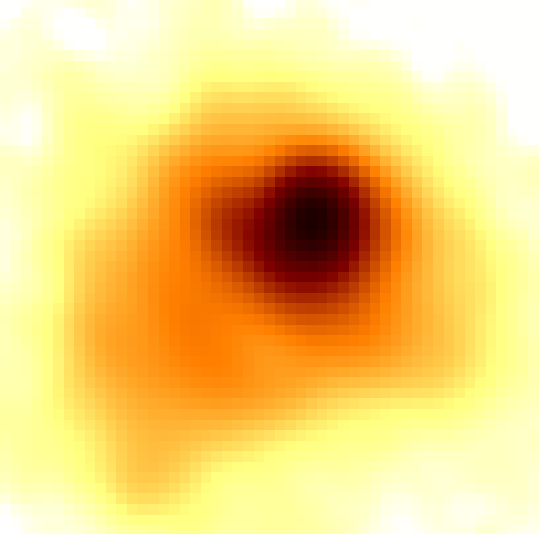}
 \hspace{0.1cm}
 \includegraphics[width=5cm,height=5cm]{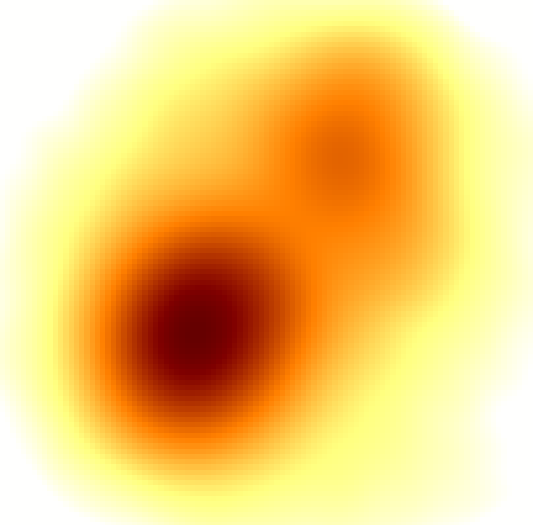}

 \vspace{0.3cm}
  
 \includegraphics[width=0.5cm,height=5cm]{images_2/lya_title.png}
 \hspace{0.1cm} 
 \includegraphics[width=5cm,height=5cm]{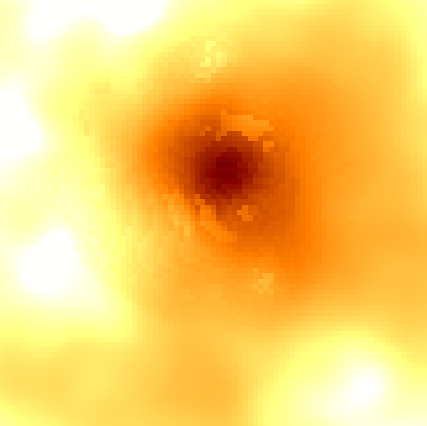}
 \hspace{0.1cm}
 \includegraphics[width=5cm,height=5cm]{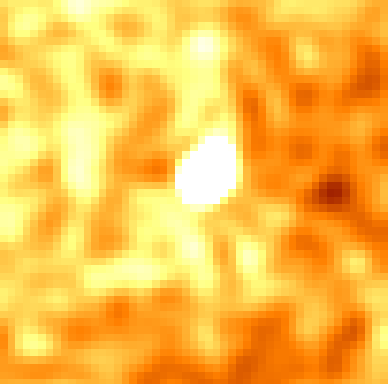}
 \hspace{0.1cm}
 \includegraphics[width=5cm,height=5cm]{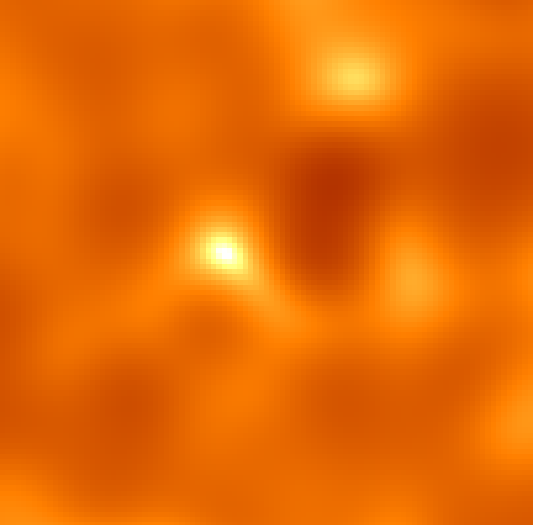}

 \vspace{0.3cm}
  
 \includegraphics[width=0.5cm,height=5cm]{images_2/ebv_title.png}
 \hspace{0.1cm}
 \includegraphics[width=5cm,height=5cm]{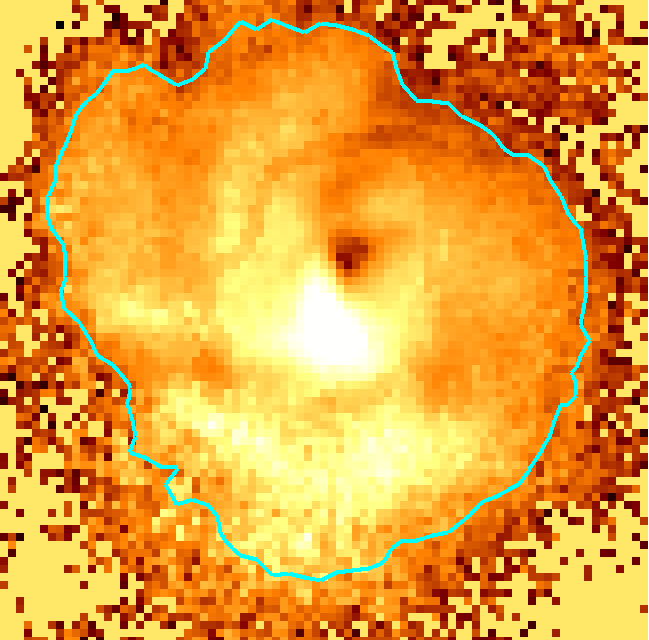}
 \hspace{0.1cm}
 \includegraphics[width=5cm,height=5cm]{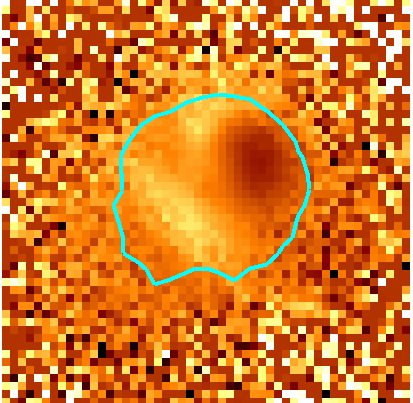}
 \hspace{0.1cm}
 \includegraphics[width=5cm,height=5cm]{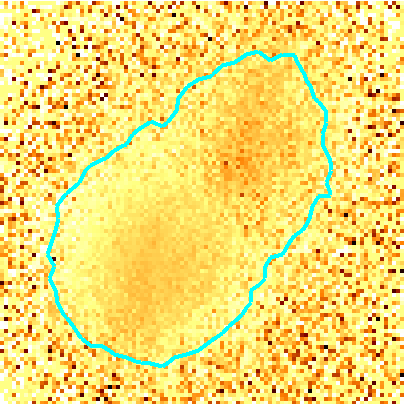}
  
 \vspace{0.3cm}

 \includegraphics[width=0.5cm,height=5cm]{images_2/lya_fuv_title.png}
 \hspace{0.1cm}
 \includegraphics[width=5cm,height=5cm]{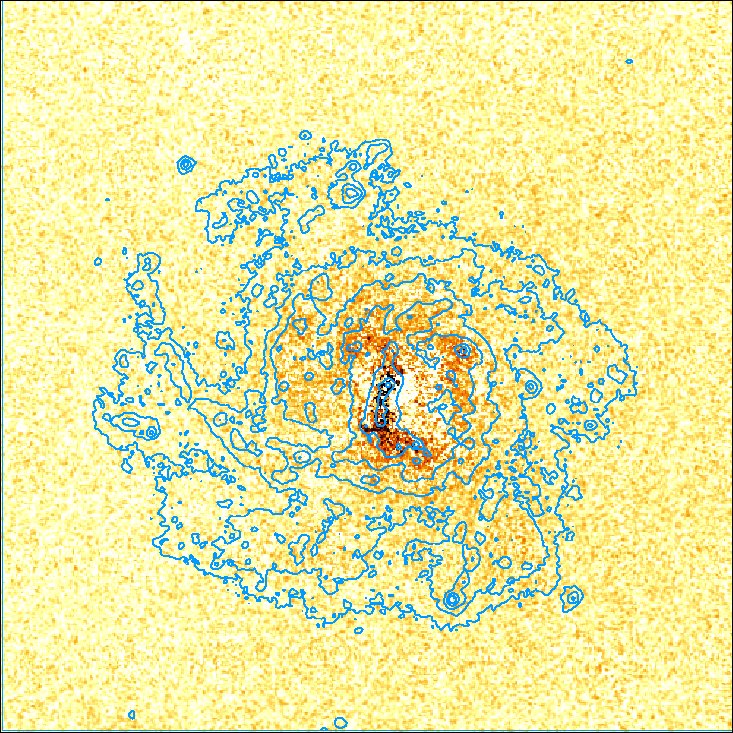}
 \hspace{0.1cm}
 \includegraphics[width=5cm,height=5cm]{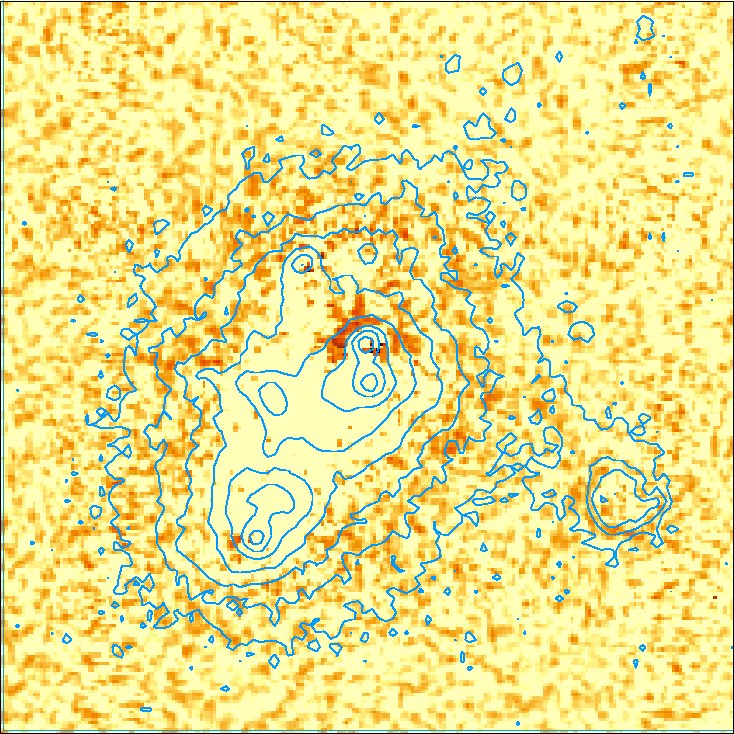}
 \hspace{0.1cm}
 \includegraphics[width=5cm,height=5cm]{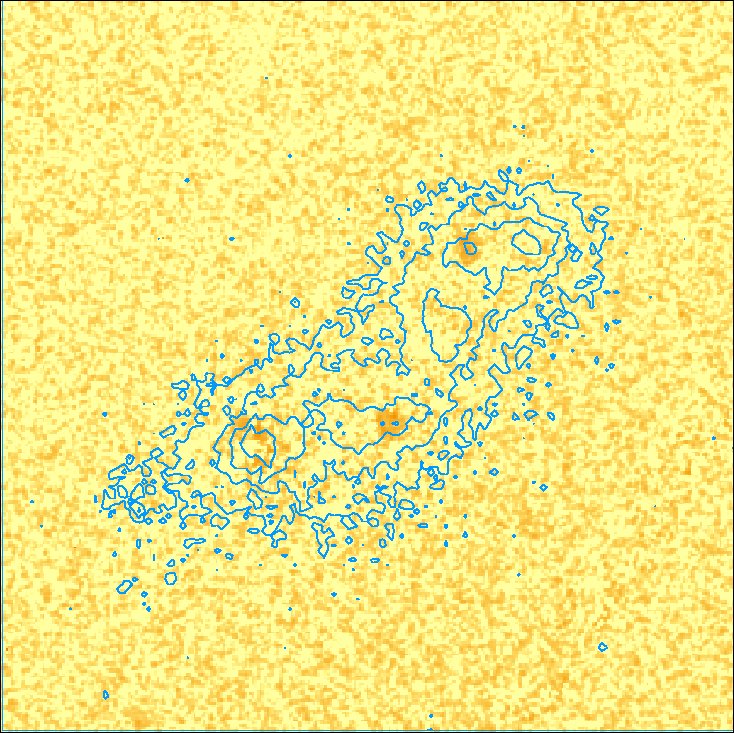}

 \vspace{0.5cm}

 \caption{Galaxy imaging (Part 2): \textit{From top to bottom:} H$\alpha$, Ly$\alpha$, E(B-V)$_{gas}$ map, and Ly$\alpha$\ as seen by HST overlayed with FUV contours. Inverted logarithmic scale is used, showing emission in black and absorption in white. The extinction map is overlayed with the mask generated following the description in Sect. \ref{discussion_sec}. Regions out of the contour are excluded from our study. The dustiest regions are in black. \textit{from left to right} with the FoV in parentheses: Ground-based images (top three rows): IRAS 08339+6517  (15\arcsec, 0.38 kpc/\arcsec), SBS 335-052 (4\arcsec, 0.27 kpc/\arcsec), Tololo 65 (8\arcsec, 0.18 kpc/\arcsec); HST images (last row): IRAS 08339+6517  (14\arcsec), SBS 335-052 (5\arcsec), Tololo 65 (8\arcsec). North is up and East to the left.} 
  \vspace{0.4cm}
 \label{images_2}
 \end{figure*}

\paragraph{\textit{\object{HARO 11}~ --}}

 The H$\alpha$\ image in Fig. \ref{images_1} in the first column shows a complex morphology with three main star-forming condensations (Kunth et al. 2003). The continuum subtracted Ly$\alpha$\ image does not delineate this morphology, showing Ly$\alpha$\ in emission in only knot C, whereas it is seen in absorption in knot A and B. The decoupling of Ly$\alpha$\ from continuum is clearly observable in the bottom frame, which represents Ly$\alpha$\ map at HST resolution overlayed by FUV (1500\AA) contours. The emission exhibits two different components consisting of a central bright knot and a low surface brightness diffuse emission. It seems that, examining the extinction map, the diffuse component is not quite regulated by the dust amount. Moreover, the bright Ly$\alpha$\ emission in knot C, corresponds to a high extinction region.

This galaxy is a well known Ly$\alpha$\ emitter \citep{kunth98}, while the detection of Lyman continuum leakage by \citet{bergvall06} is still controversial \citep{grimes07}. It has been studied in more details in \citet{hayes07a}.

\paragraph{\textit{ESO 338-04~ --}}
In the second column of Fig. \ref{images_1}, the Ly$\alpha$ image shows three main absorption regions and a surrounding bright emission. The absorption sites correspond to relatively dusty regions of the galaxy seen in the extinction map, which traps Ly$\alpha$\ photons, while the emission is not correlated with the dust content. The last component is, again, the diffuse emission around and overall the galaxy, with a low surface brightness, which is showing up the resonant decoupling of Ly$\alpha$\ photons.  Many dust features are clearly visible in the E(B-V) map, with a clumpy-like structure, and  following roughly the H$\alpha$\ structure. 
 
The Ly$\alpha$\ image is produced by matching the HST/ACS image to the NTT resolution. This results in spreading out the central absorption to more extended region, revealing small new absorption features around and dimming the surface brightness of the emission component. The bottom frame shows again how Ly$\alpha$\ is uncorrelated with the FUV continuum which traces unobscured star formation sites.

\paragraph{\textit{NGC 6090~ --}}

Third column in Fig. \ref{images_1} shows that the interacting system NGC 6090 exhibits Ly$\alpha$\ emission from each component, at a distance of about 6\arcsec\ from each other. The emission is peaking around low extinction regions, and the overlapping region between the two components, where no emission is seen appears very dusty. The extinction map of NGC 6090 shows dust pattern like a spiral structure and no evident correlation with the ionised gas traced by the H$\alpha$\ emission is seen.  

The main H$\alpha$\ structures correspond to the Ly$\alpha$\ emission components, although the largest Ly$\alpha$\ emission represents a small region in H$\alpha$, and vice-versa. This discrepancy may be due the large amount of dust in the upper left component that could destroy a part of Ly$\alpha$\ photons. Knot A appears also dustier ($E_{B-V,\mathrm{gas}}$\ $\sim$ 0.75) than knot B ($E_{B-V,\mathrm{gas}}$\ $\sim$ 0.55).

\paragraph{\textit{IRAS 08339+6517 ~ --}}
This nuclear starburst shows a spiral structure conspicuous on the H$\alpha$\ (Fig \ref{images_2}) and FUV continuum images (the latter is not shown here). However the Ly$\alpha$\ image does not resemble any of them, showing a central bright emission and the ubiquitous halo component. The FUV contours exhibit much more details in the arms of the galaxy consisting in many star clusters where Ly$\alpha$\ is absent.

The extinction map indicates a dust-free central spot. The dust  distribution has no clear relationship with the emission maps.

\paragraph{\textit{SBS 335-052~ --}}
The dust distribution is apparently correlated with the H$\alpha$\ emission in this galaxy. In Fig. \ref{images_2}, the brightest region in H$\alpha$\ presents the most important amount of dust. Are also visible in both images a shell structure and a dust-free region toward the S-E of the bright region. 

The bright dusty spot corresponds to a relatively high Ly$\alpha$\ absorption region. The galaxy shows Ly$\alpha$\ only in absorption surrounded by a faint diffuse emission. It appears that, despite a bright H$\alpha$\ or FUV continuum emission, no Ly$\alpha$\ photons manage to escape directly without scattering on neutral hydrogen.

\paragraph{\textit{Tololo 65~ --}}

In the last column of Fig. \ref{images_2}, no Ly$\alpha$\ structure can be seen for this galaxy, though the HST image shows some emission features (bottom frame), they have been smoothed to the NTT resolution. Degrading the resolution of the HST image leads to ghost features since the emission becomes as weak as the background level and is swallowed up. These artefacts in the background, although less dramatical, are also present in the other galaxies. The H$\alpha$\ map presents a two-components emission that resembles the structure of the dust content in the E(B-V) map.


\begin{figure*}[htbp]
 \centering 
 \includegraphics[width=8cm,height=6cm]{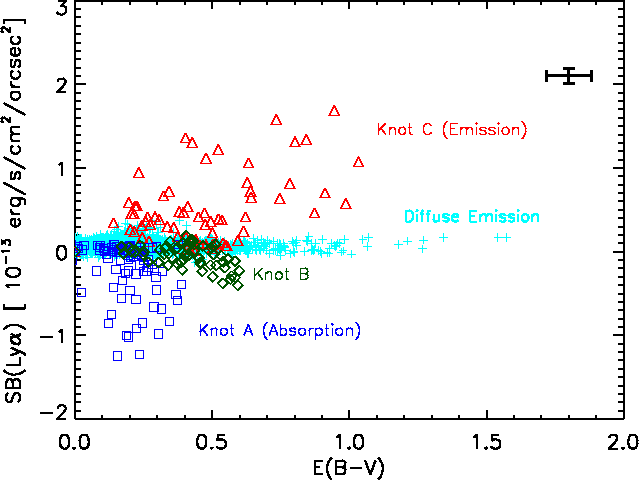}
 \hspace{0.1cm}
 \includegraphics[width=8cm,height=6cm]{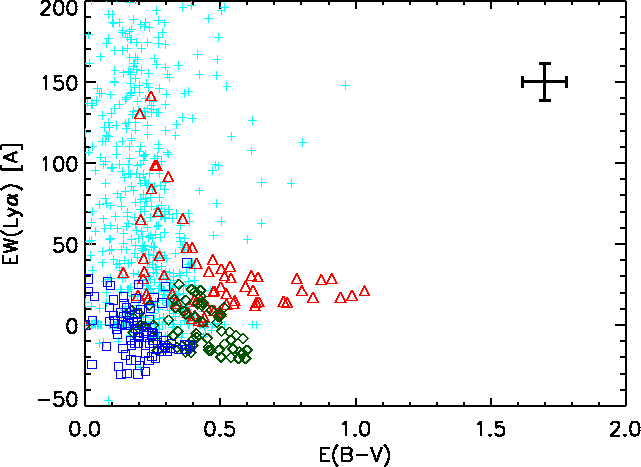}
 \vspace{0.3cm}
 \includegraphics[width=8cm,height=6cm]{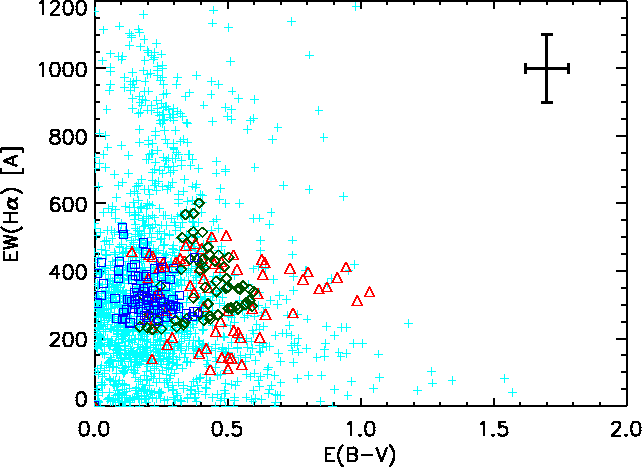}
 \hspace{0.1cm}
 \includegraphics[width=8cm,height=6cm]{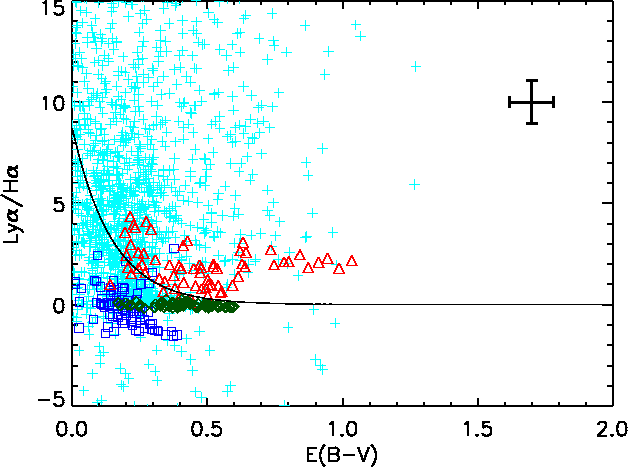}
 \caption{Haro 11 scatter plots. \textit{Top-left}: Ly$\alpha$\ surface brightness as a function of the extinction determined from the Balmer decrement. Regions of knots A, B, C and the diffuse emission are marked in the figure and represented with different colors and different symbols: Emission is in red, absorption is in blue (knot A) and green (knot B), and the diffuse emission component is in cyan. For the remaining plots, the same color code is used for the respective regions. \textit{Top-right}: Ly$\alpha$\ equivalent width \textit{vs} E(B-V). \textit{Bottom-left}: H$\alpha$\ equivalent width \textit{vs} E(B-V). \textit{Bottom-right}: Ly$\alpha$/H$\alpha$\ ratio \textit{vs} E(B-V). Overplotted on all figures the error bars corresponding to the uncertainties estimated in Sec. \ref{error_sec}} 
 \label{haro_plots}
 \end{figure*}

\paragraph{}
In order to investigate the Ly$\alpha$\ emission variations and possible correlations with  dust or other parameters at the smallest possible  scale, we have produced scatter plots from the images. In extinction maps obtained in Fig. \ref{images_1} each point represents one pixel in the galaxy region which has been isolated by masking the background at 5$\sigma$ level. Regions of interest, such as star-forming knots and emission or absorption features, are picked up using circular apertures and represented with different colors and symbols on the figures. 

\subsection{Blended emission and absorption systems}

\begin{figure}[!htbp]
 \centering 
 \includegraphics[width=8cm,height=7cm]{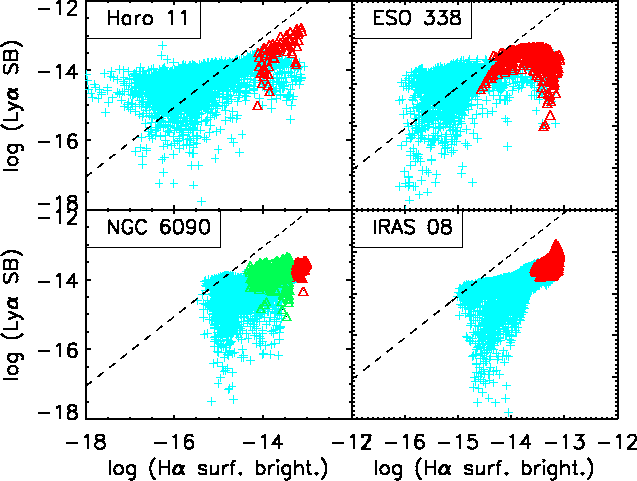}
 \caption{Pixel distribution of the Ly$\alpha$\ against H$\alpha$\ surface brightness. A logarithmic scale is used and hence shows only positive contribution (emission). The dashed line represents the case B recombination ratio. The galaxy name is given in each plot.} 
 \label{rec_plots}
 \end{figure}

 \begin{figure*}[htbp]
 \centering 
 \includegraphics[width=8cm,height=6cm]{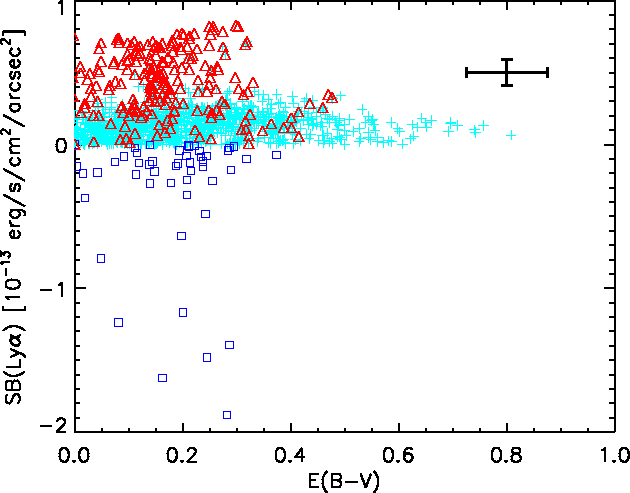}
 \hspace{0.1cm}
 \includegraphics[width=8cm,height=6cm]{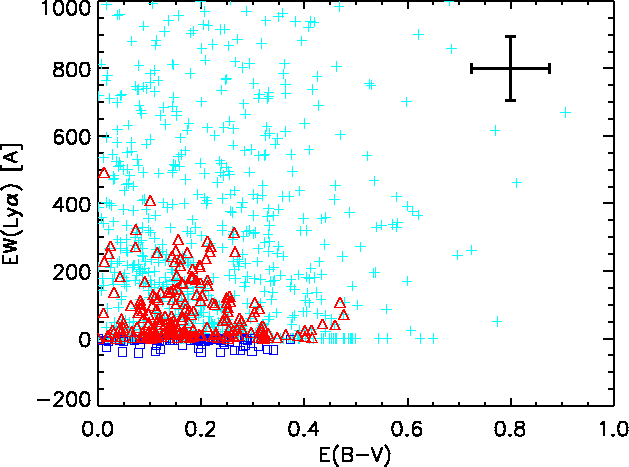}
 \vspace{0.3cm}
 \includegraphics[width=8cm,height=6cm]{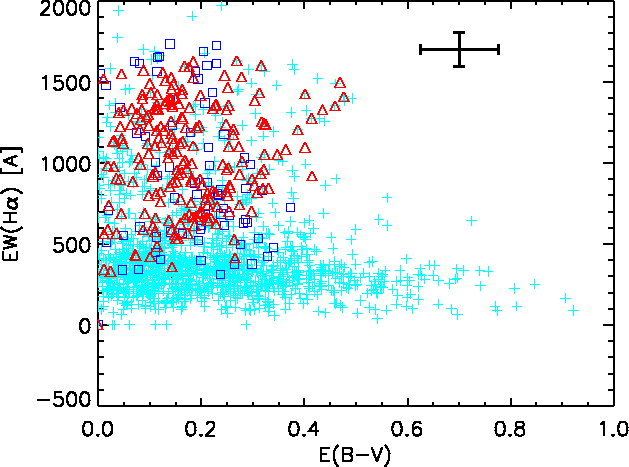}
 \hspace{0.1cm}
 \includegraphics[width=8cm,height=6cm]{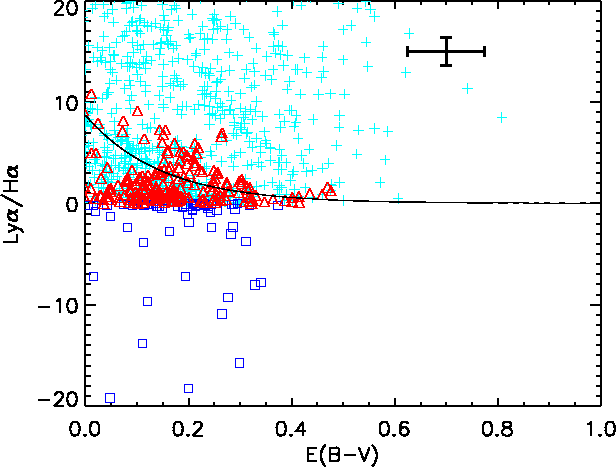}
 \caption{ESO 338-04 scatter plots. \textit{Top-left}: Ly$\alpha$\ surface brightness \textit{vs} E(B-V). Emission from the central region is in red, Absorption in the central region (knot A) is represented in blue (knot A), the surrounding emission is in red, and the diffuse emission component is in cyan. Same color code applied in all the plots. \textit{Top-right}: Ly$\alpha$\ equivalent width \textit{vs} E(B-V). \textit{Bottom-left}: H$\alpha$\ equivalent width \textit{vs} E(B-V). \textit{Bottom-right}: Ly$\alpha$/H$\alpha$\ ratio \textit{vs} E(B-V). Error bars correspond to the uncertainties estimated in Sect. \ref{error_sec}} 
 \label{eso_plots}
 \end{figure*}


\begin{figure*}[htbp]
 \centering 
 \includegraphics[width=8cm,height=6cm]{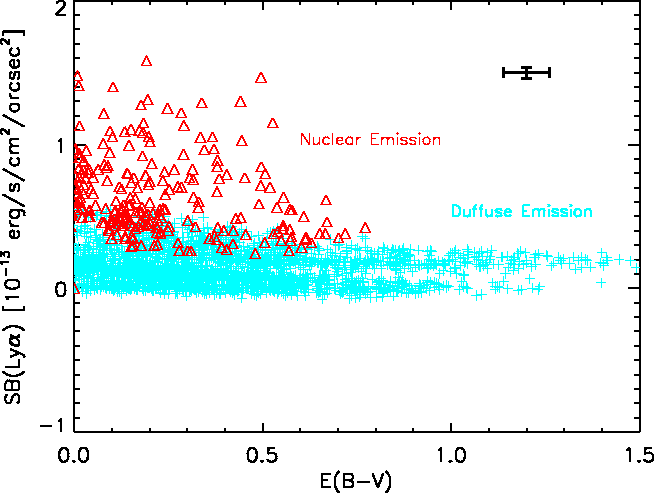}
 \hspace{0.1cm}
 \includegraphics[width=8cm,height=6cm]{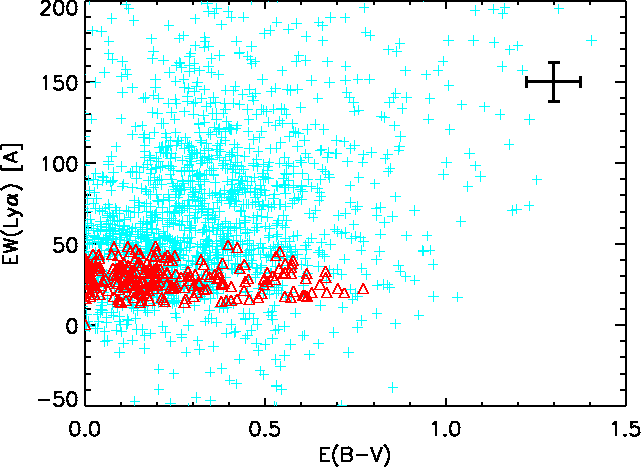}
 \vspace{0.3cm}
 \includegraphics[width=8cm,height=6cm]{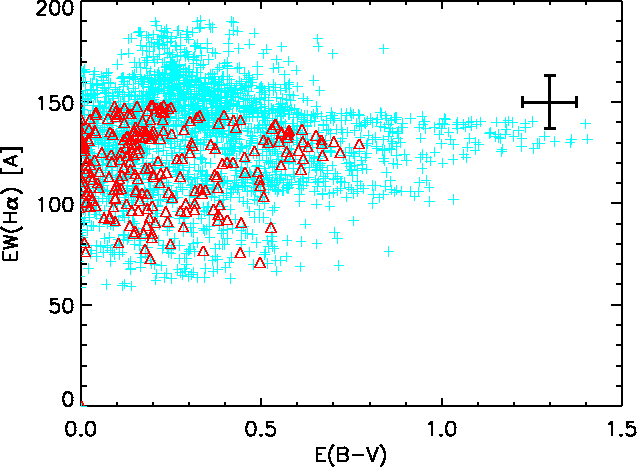}
 \hspace{0.1cm}
 \includegraphics[width=8cm,height=6cm]{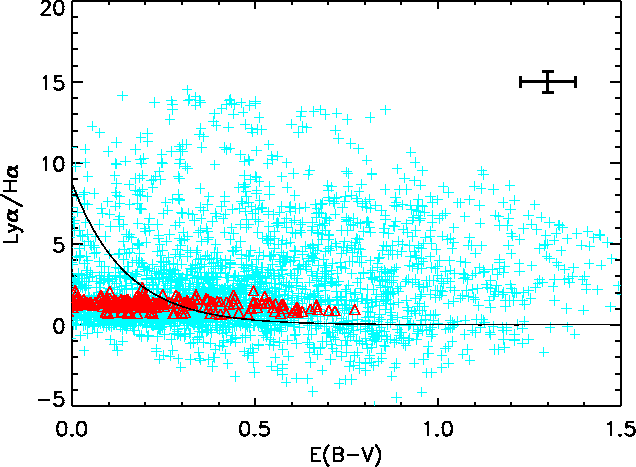}
 \caption{IRAS 08339+6517  scatter plots. \textit{Top-left}: Ly$\alpha$\ surface brightness \textit{vs} E(B-V). The nuclear Ly$\alpha$\ emission component is represented in red, and the diffuse emission component is in cyan. Same color code applied in all the plots. \textit{Top-right}: Ly$\alpha$\ equivalent width \textit{vs} E(B-V). \textit{Bottom-left}: H$\alpha$\ equivalent width \textit{vs} E(B-V). \textit{Bottom-right}: Ly$\alpha$/H$\alpha$\ ratio \textit{vs} E(B-V). Error bars correspond to the uncertainties estimated in Sec. \ref{error_sec}.} 
 \label{iras_plots}
 \end{figure*}

\paragraph{\textit{Haro 11}~ --}

The first plot in Fig. \ref{haro_plots} presents the correlation between the Ly$\alpha$\ emission and the extinction determined from the Balmer decrement tracing the dust in the gas phase ($E_{B-V,\mathrm{gas}}$). The color-code represents different regions of interest consisting of circular apertures centered on the three main knots of the galaxy, which are marked on the Ly$\alpha$\ image. We can see a diffuse emission component extending up to $E_{B-V}$\ $\sim$ 1.5. Knot $C$ shows a bright and spread emission with a mean extinction of 0.48,  while the absorption is essentially localised around knot $A$ and $B$. The mean extinctions in each knot are derived using the ratio of integrated H$\alpha$\ and H$\beta$\ fluxes within the corresponding apertures. 

The presence of two distinct emission components is indicative of two different physical processes implied in the escape of Ly$\alpha$\ photons. Firstly the diffuse component that shows the resonant decoupling of Ly$\alpha$\ photons resonantly scattered on hydrogen atoms until they escape far away from there production sites (thus, experiencing a large range of extinction). On the other hand, the emission from knot $C$ is less spread and represents photons escaping directly from this small region with a mean extinction $E_{B-V}$\ $\sim$ 0.48. The Ly$\alpha$\ resonant decoupling is also visible in Fig. \ref{rec_plots} representing the correlation between the Ly$\alpha$\ and the H$\alpha$\ fluxes in log-scale, hence, showing only positive pixels (in emission). We observe a first component at a low and roughly constant Ly$\alpha$\ surface brightness (around $10^{-14}$ erg~s$^{-1}$~cm$^{-2}$~arcsec$^{-2}$), independent of H$\alpha$\ emission. Thanks to their resonant scattering, Ly$\alpha$\ photons reach regions where non-resonant photons, such as H$\alpha$, are absent, making the Ly$\alpha$/H$\alpha$\ ratio higher than the Case B level (represented by a dashed line in the figure). The second component at higher Ly$\alpha$\ and H$\alpha$\ fluxes ($f_{\mathrm{Ly}\alpha}$\ $\ge 10^{-14}$ erg~s$^{-1}$~cm$^{-2}$~arcsec$^{-2}$ ) is always below the predicted recombination value. These pixels represent regions where Ly$\alpha$\ photons escape directly from their production site where H$\alpha$\ is also produced. 

That we see Ly$\alpha$\ in emission from knot C with $E_{B-V}$\ $\sim$ 0.48 whereas absorption is seen in knots A and B with $E_{B-V}$\ $\sim$ 0.2 and 0.41 respectively is interesting. Indeed Ly$\alpha$\ photons manage to escape from regions with higher extinction than those of a pure absorption. The dust content is clearly not the main driver in the escape process of Ly$\alpha$\ photons. A great covering of static H{\sc i}\ column density in knot A and an expanding neutral ISM and/or ionised H{\sc ii}\ holes in knot C may lead to this observation. 

Looking at Ly$\alpha$\ equivalent width, we observe that the diffuse emission shows relatively high $EW_{\mathrm{Ly}\alpha}$\ whereas the pure emission equivalent width in knot C (in red) is much weaker. This suggests that a hard FUV radiation could create ionised holes through which Ly$\alpha$\ photons may escape, in a inhomogeneous distribution of H{\sc i}\ and dust. In this case of multi-phase ISM, it has been shown \citep{neufeld91, hansen06} that, thanks to their scattering on the dusty H{\sc i}\ clumps, Ly$\alpha$\  escapes in an easier way than non-resonant photons. We also observe that the diffuse emission (in cyan) corresponds to the highest equivalent width observed ($EW_{\mathrm{Ly}\alpha}$\ higher than 200 $\AA$), since it represents photons that have scattered far away from their production sites and escape where the Ly$\alpha$\ continuum is lower. This decline in the emission is, again, symptomatic of the resonant nature of Ly$\alpha$\ photons. Indeed, when we plot the equivalent width of H$\alpha$\ against the extinction, we do not see any correlation, as we expect for non-resonant lines, since the online and the continuum photons are regulated by the same physical processes according to the dust content. The last figure shows how the Ly$\alpha$/H$\alpha$\ ratio evolves according to the amount of dust. In a classical view, considering only the selective extinction at the two wavelengths, and a case B intrinsic ratio of 8.7 \citep{brocklehurst71}, we expect to have an exponential decline represented by the dark curve. The resonant nature of the Ly$\alpha$\ photons leads to a different result. We observe a high dispersion for the halo component and an emission from knot C above the predicted level at higher extinction, which might support the view of a scattering in inhomogeneous ionised ISM that favors preferentially the escape of Ly$\alpha$\ photons.

 \begin{figure*}[!htbp]
 \centering 
 \includegraphics[width=8cm,height=6cm]{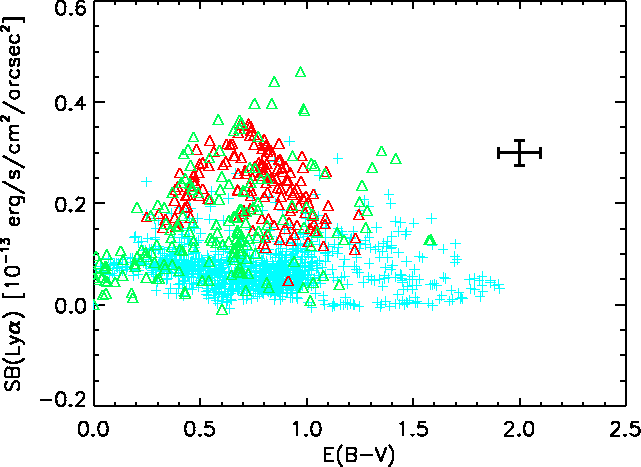}
 \hspace{0.1cm}
 \includegraphics[width=8cm,height=6cm]{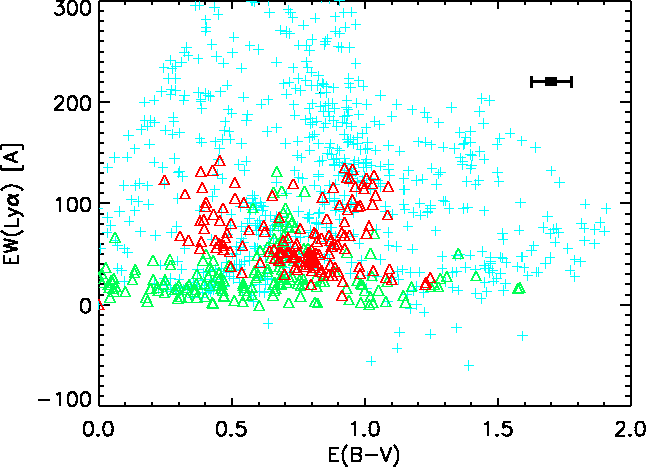}
 \vspace{0.3cm}
 \includegraphics[width=8cm,height=6cm]{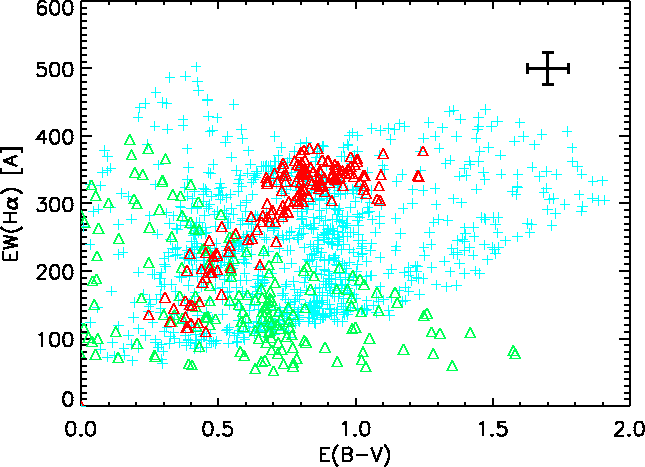}
 \hspace{0.1cm}
 \includegraphics[width=8cm,height=6cm]{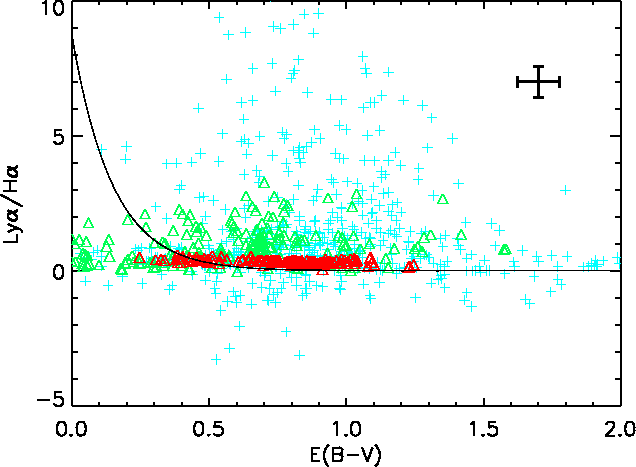}
 \caption{NGC 6090 scatter plots. \textit{Top-left}: Ly$\alpha$\ surface brightness \textit{vs} E(B-V). Emission from knot A is in red, and from knot B is in green. Diffuse emission component is represented as usual in cyan. Same color code applied in all the plots. \textit{Top-right}: Ly$\alpha$\ equivalent width \textit{vs} E(B-V). \textit{Bottom-left}: H$\alpha$\ equivalent width \textit{vs} E(B-V). \textit{Bottom-right}: Ly$\alpha$/H$\alpha$\ ratio \textit{vs} E(B-V). Error bars correspond to the uncertainties estimated in Sec. \ref{error_sec}.} 
 \label{ngc_plots}
 \end{figure*}

\paragraph{\textit{ESO 338-04}~ --}

This galaxy shows a similar pixel distribution in Ly$\alpha$-$E_{B-V}$\ space to that of Haro 11 (Fig. \ref{eso_plots}). We see both emission and pure absorption features at, approximately, the same extinction (mean $E_{B-V}$\ $\sim$ 0.22 and 0.23 for absorption and global emission respectively). The different regions are marked in the scatter plots with different colors consisting of a central absorption (in blue) surrounded by emission features (red). We also observe a halo of diffuse emission surrounding the starburst regions and independent of the extinction. However the diffuse emission does not exceed 5 $\times$ 10$^{-14}$ erg~s$^{-1}$~cm$^{-2}$~arcsec$^{-2}$, it accounts for about 70\% of the total Ly$\alpha$\ emission from the galaxy. The same value was found by \citet{hayes05}, although a different masking was used.

The absorption is seen only in the knot A \citep[following the nomenclature of][]{hayes05} superimposed on a region with an average extinction of $E_{B-V}$\ $\sim$ 0.2 (mean extinction calculated, as for Haro 11, using integrated H$\alpha$\ and H$\beta$\ fluxes over this knot). \citet{ostlin03} from longslit spectroscopy measurement of H$\alpha$/H$\beta$\ (the slit being positioned on the East-West direction) found E(B-V) values within a range comparable to our result ($E_{B-V,\mathrm{gas}}$\ $\sim$ 0 - 0.25), except toward the east edge of the galaxy where a dustier region is seen in our extinction map ($E_{B-V,\mathrm{gas}}$\ $\sim$ 0.4). Same extinction as in knot A is observed in its surrounding regions, where we found Ly$\alpha$\ in emission. There is a trend of decline in the emission (though, with high dispersion) (until $E_{B-V}$\ $\sim$ 0.6). The radiative transfer of Ly$\alpha$\ photons in a static (or nearly static) ISM may lead to the present situation: In the central region (knot A), Ly$\alpha$\ photons are produced and immediately absorbed by neutral hydrogen. They are re-emitted according to space and frequency redistribution probability. Eventually, Ly$\alpha$\ photons resonantly scatter in the wings until they reach a frequency which is far enough from the line center (where the absorption probability is close to unity) to be able to escape from the neutral medium. This also leads to a diffusion in space which explains the emission seen around the central absorption at the same extinction. Actually, this effect is also seen in Haro 11 with the HST resolution (Fig. \ref{images_1}), but does not appear at the NTT resolution due to the smoothing effects. It seems that, as shown by Ly$\alpha$\ radiative transfer models \citep[e.g.][]{verhamme06}, Ly$\alpha$\ photons do not easily escape  from where they have been produced but scatter in space and frequency avoiding absorption by HI atoms at line center ($\nu_{0}$ = 2.46 $\times$ 10$^{15}$ Hz).

We note in Fig. \ref{eso_plots} very high Ly$\alpha$\ equivalent width, ranging from -10 \AA\ for the absorption (blue) to $\sim$250 \AA\ in the surrounding emission (red). Such high EW might be due either to the resonant scattering mechanism that allows Ly$\alpha$\ photons to travel where continuum photons can not, or to a multi-phase configuration of the ISM where, \citet{hansen06} noted for reasonable HI column density and dust amount, that continuum is preferentially extinguished, boosting the initial EW easily by a factor of 2-5. This configuration may also explain the high $EW_{\mathrm{Ly}\alpha}$\ observed in some high-redshift galaxies, extending up to 150 $\AA$ in LALA $ z \sim$ 5.7 sources \citep{rhoads03} in the case of spectroscopically confirmed candidates (higher possible equivalent widths were found in the imaging survey). On the other hand, H$\alpha$\ equivalent width map shows similar distribution to that of Haro 11 without any correlation. 

The last frame shows the evolution of Ly$\alpha$/H$\alpha$\ ratio with the extinction. In the central region around knot A, Ly$\alpha$/H$\alpha$\ follows loosely the theoretical curve (marked in black).

However, the absorption region does not show any trend of decrease in Ly$\alpha$/H$\alpha$\ ratio. Rather, we see a dispersion, as seen in the Ly$\alpha$-$E_{B-V}$\ plot, ranging from $E_{B-V}$\ $\sim$ 0 to 0.5. 

 \begin{figure*}[!htbp]
 \centering 
 \includegraphics[width=8cm,height=6cm]{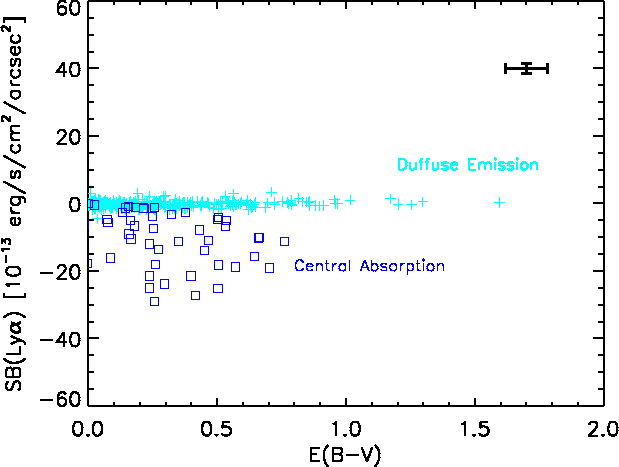}
 \hspace{0.1cm}
 \includegraphics[width=8cm,height=6cm]{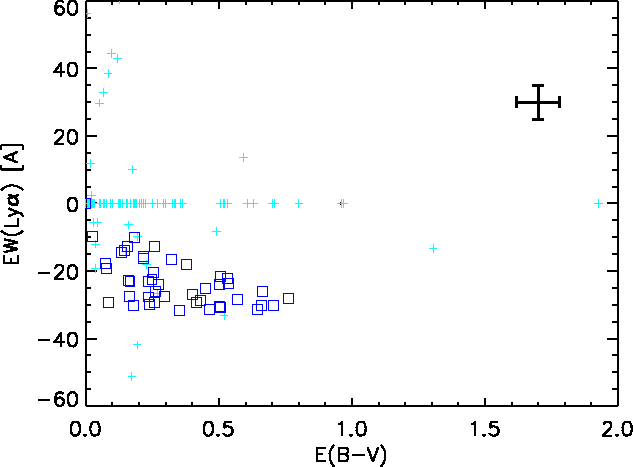}
 \vspace{0.3cm}
 \includegraphics[width=8cm,height=6cm]{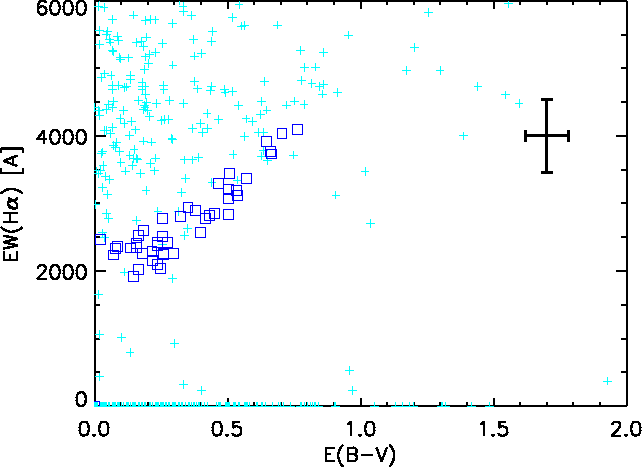}
 \hspace{0.1cm}
 \includegraphics[width=8cm,height=6cm]{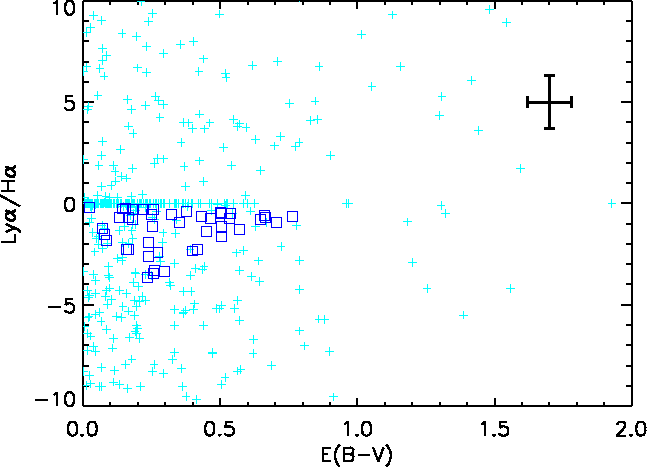}
 \caption{SBS 335-052 scatter plots. \textit{Top-left}: Ly$\alpha$\ surface brightness \textit{vs} E(B-V). The damped central absorption is represented in red, and the diffuse emission component is in cyan. Same color code applied in all the plots. \textit{Top-right}: Ly$\alpha$\ equivalent width \textit{vs} E(B-V). \textit{Bottom-left}: H$\alpha$\ equivalent width \textit{vs} E(B-V). \textit{Bottom-right}: Ly$\alpha$/H$\alpha$\ ratio \textit{vs} E(B-V). Error bars correspond to the uncertainties described in Sec. \ref{error_sec}.} 
 \label{sbs_plots}
 \end{figure*}

\subsection{Emission systems: NGC 6090 \& IRAS 08339+6517 }
\paragraph{}

Both of these galaxies exhibit Ly$\alpha$\ only in emission, with little signs of direct absorption, as we have seen on Fig. \ref{images_1} and \ref{images_2}. As usual, the first plot in Fig. \ref{iras_plots} and \ref{ngc_plots} shows Ly$\alpha$\ flux function of the color excess $E_{B-V}$. In both systems, we do see the diffuse component independent of $E_{B-V}$\ at low Ly$\alpha$\ flux. 

Although, the calibration procedure used for IRAS 08339+6517  data is not accurate (see section \ref{ground_obs_sec}), the average extinction found for this galaxy ( $E_{B-V}$\ $\sim$ 0.12 ) is still close to the value estimated by \citet{gonzalez98}. 
 
STIS spectroscopy \citep{mashesse03} revealed a P-Cygni profile, with a red wing shifted by $\sim 300$ km~s$^{-1}$ with respect to the H{\sc ii~} region velocity determined from the optical emission lines. The extension of the neutral gas shell is  found to be very large, with a diameter of around 10 kpc, which is larger than the aperture used here to isolate the emission region in the center of the galaxy ($\sim$ 2 kpc). Thus, the central emission (red component)  is observed through an expanding shell approaching us at a velocity of $\sim 300$ km~s$^{-1}$. 

Same Ly$\alpha$\ profile has been observed for \textit{NGC 6090} by \citet{gonzalez98}, with a velocity offset with respect to blueshifted interstellar absorption lines, indicative of large scale high-velocity outflows of gas, of several hundred km~s$^{-1}$. We differentiate two emission blobs
(red for A and green for B in Fig. \ref{ngc_plots}). Component A appears at a mean E(B-V) around 0.74 whereas the second one at E(B-V) $\sim$ 0.54. The diffuse emission, also in this galaxy, is the only component that attains the recombination level (Figure \ref{rec_plots}) while its bulk stays quite below.

Again, regarding these two galaxies, the ISM kinematics must play an important role in the escape of Ly$\alpha$\ photons through neutral gas shells, making Ly$\alpha$\ \textit{less sensitive} to the dust content. It leads to the observed dispersion in the emission over a large range of extinction for these two galaxies. To confirm this visual inspection, we have performed a Spearman's statistical test which gives the probability that a correlation between Ly$\alpha$\ and $E_{B-V}$ exists. The Spearman's correlation coefficient $\rho$\ can take values from -1 to +1. A value of +1 shows that the variables are perfectly correlated with a positive slope, a value of -1 shows that the variables are anti-correlated (negative slope), and a value of  0 shows that the variables are completely independent. While we would expect an anti-correlation between these two variables, we found $\rho \sim$ 0.15 for both galaxies (whereas $\rho \sim$ -0.2 and -0.05 for Haro 11 and ESO 338-04 respectively), confirming that Ly$\alpha$\ is less sensitive to dust in these systems. For SBS 335-052 we found $\rho \sim$ -0.56.

The Ly$\alpha$\ equivalent width appears to be not correlated with the dust in IRAS 08339+6517 . This absence of correlation is also seen when observing the evolution of the line ratio Ly$\alpha$/H$\alpha$\ which is around 2 and stays constant as the extinction increases supporting the importance of the ISM kinematics.

The situation is slightly different for NGC 6090, where in knot A we see values more clearly above the theoretical Ly$\alpha$/H$\alpha$\ curve (as seen for Haro 11). In addition to kinematics, we may be in presence of clumpy ISM configuration as well (as it is suggested by the images), that allows Ly$\alpha$\ to get out through the ionised inter-clouds medium. Furthermore, the two knots show different equivalent widths: $EW_{\mathrm{Ly}\alpha}$\ $\sim$ 54 \AA\ in knot A and 20 \AA\ in knot B. Note that, the diffuse emission presents higher values because of the numerous resonant scatterings that have experienced these photons. This is still valid for Ly$\alpha$/H$\alpha$\ ratio.

\subsection{Damped absorption system: SBS 335-052}
\label{absorption_sec}

HST/GHRS spectroscopy \citep{thuan97a} has revealed these metal-deficient BCDs to be damped Ly$\alpha$\ absorbers. They don't show, indeed in our images, any direct emission but a low surface brightness diffuse halo. The Ly$\alpha$\ photons  manage to escape from these galaxies after multiple scattering events. Figure \ref{sbs_plots} shows how Ly$\alpha$\ is related to different physical parameters. We see, for \textit{SBS 335-052}, in the first frame, a weak diffuse component which accounts for the whole Ly$\alpha$\ emission. In a classical vision, where dust is the main parameter responsible of the destruction of Ly$\alpha$\ photons, we expect a declining relationship between Ly$\alpha$\ flux and the dust extinction. It is precisely what we observe in this galaxy. A strong absorption ($f_{\mathrm{Ly}\alpha}$\ $\sim -3 \times 10^{-12}$ erg~s$^{-1}$~cm$^{-2}$~arcsec$^{-2}$) is seen, with a weak declining trend in the range  0 \lsim $E_{B-V}$\ \lsim 0.7. This result suggests that the dust is playing, in this case, an important role in the escape of Ly$\alpha$\ photons. Very Large Array 21 cm observations \citep{thuan97b} show the BCD to be embedded in a large H{\sc i}\ cloud. The H{\sc i}\ column density in the GHRS aperture derived by these authors is very large, $N$(H{\sc i}) = 7.0 $\times$ 10$^{21}$ cm$^{-2}$. According to \citet{mashesse03} and the evolutionary models of \citet{tenorio99}, this galaxy is a very young starburst of which the age of the stellar population is too small to have ionised the whole surrounding medium, leading to a great amount of neutral gas covering the massive stars. In addition, the H{\sc i}\ cloud is  static with respect to the central H{\sc ii~} region as the 21 cm and the emission line velocities are in good agreement. As a result, the Ly$\alpha$\ photons  after a large scattering process on hydrogen atoms increase their mean path and  the probability to be absorbed by dust grains. The combination of a great H{\sc i}\ column density in front of the Ly$\alpha$\ production sites and the absence of kinematics in this neutral envelope makes Ly$\alpha$\ photons very likely to be destroyed by dust, hence turning the dust extinction into a very important Ly$\alpha$\ escape regulator. This configuration, with the related dust correlations, was observed for the knot A in Haro 11. It also appears that the Ly$\alpha$\ equivalent width is declining with  the extinction.

\textit{TOLOLO 65}, in the same way, shows very weak Ly$\alpha$\ emission through a diffuse component without direct emission. 
GHRS spectrum \citep{thuan97b} shows, with an acceptable signal-to-noise ratio although poorer than that of SBS 335-052, that this galaxy is a pure Ly$\alpha$\ absorber. This is consistent with the very faint and diffuse emission found here, which is still not significantly greater than the background level where the degraded resolution wipes out the absorption seen in the original HST/ACS image by \citet{ostlin08}.

\section{Discussion}
\label{discussion_sec}

Rather than the detailed description performed in the last section, for each galaxy to investigate  how  Ly$\alpha$\ emission is related to other parameters, especially dust, at a small scale; we discuss hereinafter the global characteristics of the galaxies based upon the integrated quantities in defined apertures. These apertures are defined for all the galaxies by masking regions that shows f(H$\beta$) below a threshold of 5 times the standard deviation of the background in the H$\beta$\ image. This study of course does not allow any meaningful statistics with only six galaxies in our hands, that have been however selected in order to sample the  largest possible starburst physical parameters. Yet we can outline some interesting trends when compared to the small-scale approach and how far some properties could be smoothed, or not, at the galaxy scale.  

\subsection{The role of the dust in Ly$\alpha$\ obscuration}
\label{role_dust_sec}
\begin{table*}[!htbp]
\centering
\renewcommand{\footnoterule}{}  
\begin{tabular}{lcccccccccc}
\hline \hline \\
Target       &Aperture& $f_{\mathrm{Ly}\alpha}$\    & $f_{\mathrm{H}\alpha}$~      & $f_{\mathrm{H}\beta}$~      & Ly$\alpha$/H$\alpha$\  &(Ly$\alpha$/H$\alpha$)$_C$\ & H$\alpha$/H$\beta$\ & EW(Ly$\alpha$) & EW(H$\alpha$)\\
         & Size   & (erg~s$^{-1}$~cm$^{-2}$)& (erg~s$^{-1}$~cm$^{-2}$) & (erg~s$^{-1}$~cm$^{-2}$) &         &          &        &   (\AA)       &  (\AA) \\
       \\
\hline \\
Haro 11      &143 (23)   & 1.3e-12 &  2.3e-12  & 5.48e-13  &  0.57    &  5.48  &  4.14     &  22.8 & 523  \\
ESO 338-IG04 &267 (10)   & 2.5e-12 &  2.5e-12  & 8.1e-13   &  0.98    &  1.69  &  3.12     &  15.8 & 479     \\
SBS 0335-052 &3.40 (0.23)& -4.0e-13 &  3.6e-13  & 9.9e-14   & -1.12    & -4.8   &  3.62   & -27  & 808      \\
NGC 6090     &147 (49)   & 6.6e-13 &  1.4e-12  & 2.1e-13   &  0.46    &  82    &  6.66   &  62   & 180     \\
IRAS 08339+6517   &157 (23)   & 3.1e-12 &  2.3e-13  & 6.9e-13   &  1.36    &  3     &  3.26   & 45.6  & 140   \\
Tololo 65    &22.7 (0.75)& 5.e-14  &  1.8e-13  & 4.9e-14   &  0.28    &  1.15  &  3.6    &  9.1 & 1153     \\
\\
\hline
\end{tabular}
\caption{ Integrated fluxes and equivalent widths for the six galaxies of the sample. The integration aperture is defined by masking regions below a certain threshold based upon H$\beta$\ flux ($f_{\mathrm{H}\beta}$~~$\geq$ 5$\sigma$, where $\sigma$ is the background standard deviation in the same line). The aperture size is given in the first column in (\arcsec)$^2$ and in kpc$^2$ in parentheses. The quantities are corrected for galactic extinction \citep{schlegel98} but not for internal reddening, except for (Ly$\alpha$/H$\alpha$)$_C$, which has been dereddened with $E_{B-V,\mathrm{gas}}$\ using \citet{cardelli89} parameterization.}
\label{catalog}
\end{table*}
\begin{figure}[!htbp]
 \centering 
 \includegraphics[width=7cm,height=6cm]{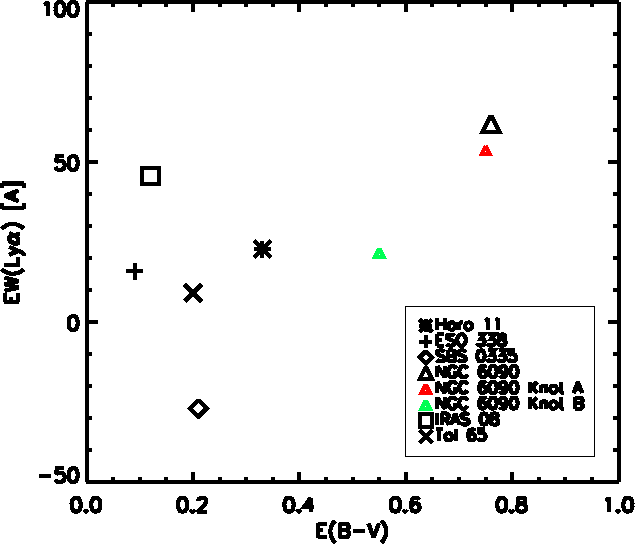}
 \caption{Scatter plot of integrated quantities, representing in each galaxy the Ly$\alpha$\ equivalent width as a function of the extinction. E(B-V) is derived from the ratio between the integrated H$\alpha$\ and H$\beta$\ fluxes. The legend of symbols assigned to the galaxies are showed in the inset. We have included in this plot the two components of NGC 6090 that we have considered separately.} 
 \label{gen_ewlya}
 \end{figure}

\paragraph{}
We show in Table \ref{catalog} the photometric properties of our sample, integrated in an aperture defined by a mask based upon a H$\beta$\ flux threshold above the sky level, in order to get the extended emission of the Balmer lines and the Ly$\alpha$\ diffuse emission. As mentioned before, all the quantities presented have been corrected for galactic extinction using \citet{schlegel98} method. 
The integration of the Ly$\alpha$\ flux within the mask aperture results in five candidates of our sample being emitters and only one net absorber (SBS 335-052). Defining this galaxy as a net absorber means that the sum of the flux overall the galaxy gives a negative result, however we do see emission in some regions. Though, the flux and the measurements derived from are very sensitive to the aperture size. We can expect that we are losing a part of the weak diffuse Ly$\alpha$\ emission and the most extended H$\alpha$\ or H$\beta$\ emission, which could be attained by deeper observations. Thus, measurements such as the escape fraction of Ly$\alpha$\ may change according to the adopted mask size since Ly$\alpha$\ can scatter further from the production sites than Balmer or continuum photons.

The Ly$\alpha$/H$\alpha$\ ratio ranges from -1.12 to 1.36, showing Ly$\alpha$\ emission much weaker for all the sample than predicted by recombination theory even, in most cases, when corrected for the differential extinction at these different wavelengths. Previous observations \citep{terlevich93, giavalisco96} yielded the same conclusions. The dust is just the final stage of the process responsible for the obscuration of Ly$\alpha$\ photons, after the resonant scattering in an homogeneous medium that increase their mean path, which makes the recombination ratio not only regulated by the dust (as seen from the H$\alpha$/H$\beta$\ ratio which traces the nebular dust). Setting the Ly$\alpha$\ equivalent width against the different parameters of Table \ref{catalog} ends up with no clear trends. 
Plotting the Ly$\alpha$\ equivalent width against E(B-V) allows us to probe the difference in extinction between resonant and non-resonant radiations, knowing that $EW_{\mathrm{Ly}\alpha}$\ is not affected by the selective extinction (i.e. independent of the dust extinction curve). We see, indeed, in Fig. \ref{gen_ewlya}, a rather scattered data points and no well-observed correlation. Since we are dealing with resonant radiation investigated from only one line-of-sight one expect that the geometry and the distribution of the dust layers around the emitting regions may affect the observed scatter of $EW_{\mathrm{Ly}\alpha}$\ according to the extinction. This relation depends also on the intrinsic EWs.

When we calculate the dereddened ratio (Ly$\alpha$/H$\alpha$)$_C$, rather than $EW_{\mathrm{Ly}\alpha}$, in Table \ref{catalog}, we observe that it lies below the recombination value (8.7), expected if the extinction is only due to dust, for five galaxies. Only NGC 6090 shows (Ly$\alpha$/H$\alpha$)$_C$\ that exceeds this recombination level. Since the resonant Ly$\alpha$\ radiation is spatially decoupled from continuum or Balmer lines, the Ly$\alpha$\ photons may suffer different extinction than traced by the Balmer decrement and lead to this overestimation of the reddening correction. In other words, the Ly$\alpha$\ photons that are detected and survived to numerous scattering and various extinctions, have been travelling through ISM regions where dust amount locally departs from  the average. Including a large part of the Ly$\alpha$\ diffuse emission in the integration aperture could, for the same reason, contribute to this effect.

We can retain from the above analysis that assuming simple dust extinction fails to recover the intrinsic Ly$\alpha$/H$\alpha$\ ratio, where the role of the dust is, in most cases, underestimated because of the resonant scattering phenomenon of Ly$\alpha$ .

\subsection{Galaxy sample and Ly$\alpha$\ emission morphology}
\label{morph_sec}

\begin{figure}[!htbp]
 \centering 
 \includegraphics[width=7cm,height=6cm]{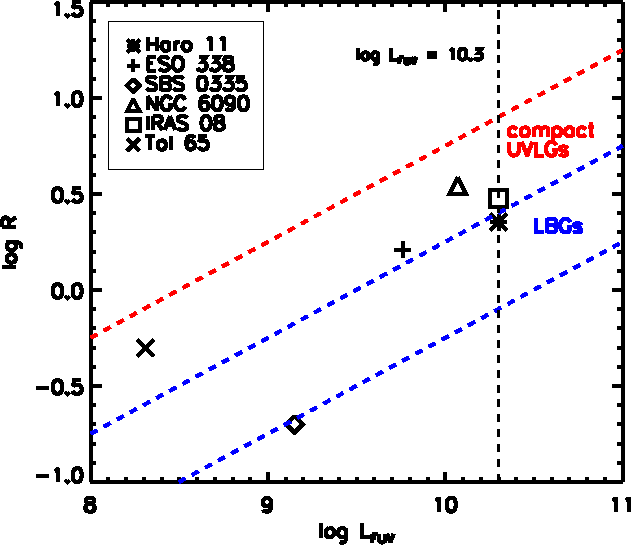}
 \caption{FUV luminosity derived from F140LP flux (1500 \AA) in units of L$_{\odot}$ versus galaxy radius (in kpc). The masking technique based on the FUV background is applied to derive integrated $L_{\rm FUV}$ and the equivalent radius $R_{UV} = \sqrt{area/\pi}$. The upper dotted line corresponds to compact-large UVLGs threshold and lower ones are LBG limits in terms of $l_{FUV}$ (see text). LBGS and UVLGs spaces are marked on the figure.} 
 \label{lfuv_r}
 \end{figure}

\begin{figure}[!htbp]
 \centering 
 \includegraphics[width=4cm,height=4cm]{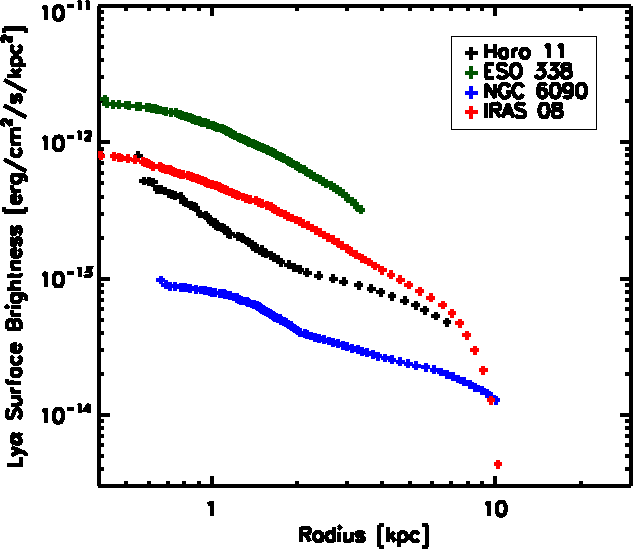}
 \includegraphics[width=4cm,height=4cm]{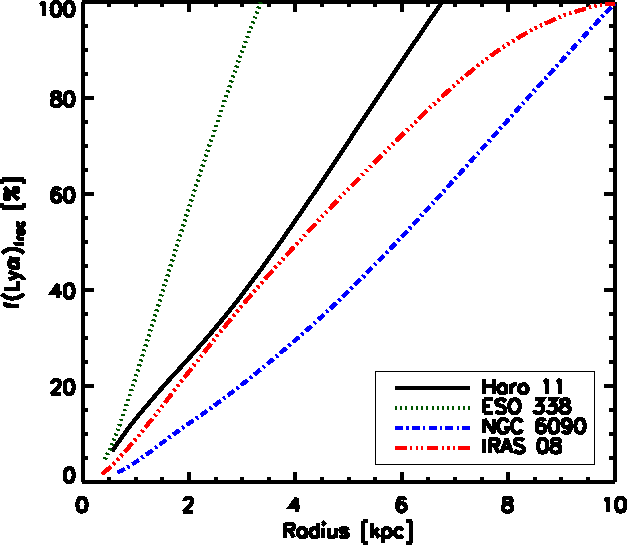}
 \caption{Ly$\alpha$\ emission morphology. The \textit{left plot} shows the Ly$\alpha$\ surface brightness integrated in radius bins against the distance from the brightest emission center. On the \textit{right plot}, we see the evolution of the Ly$\alpha$\ emission fraction cumulated in varying apertures, as a function of the equivalent radius of these apertures. It shows what fraction of Ly$\alpha$\ is emitted within a given aperture around the bright emission center. SBS 335-052 and Tololo 65, which show no direct bright but only weak diffuse emission, are not represented.} 
 \label{morph_plots}
 \end{figure}

We present here a comparison of our sample properties to their high redshift counterparts, such as Lyman Break Galaxies (LBGs). The FUV luminosity is that derived from integration within the apertures based on FUV background mask and the radius corresponds to $R_{UV} = \sqrt{area/\pi}$. The galaxies span a large range in FUV luminosity 8.3 $\leq log(L_{\rm FUV}/L_{\odot}) \leq$ 10.3 that reaches, for two galaxies, the characteristic LBGs luminosity, and are relatively compact systems, similar to LBGs radii range (log(R$_{UV}(kpc)$)$\sim$ 0-0.5). Accordingly, the FUV surface brightness $l_{FUV}$ of the sample corresponds to that observed in Ultraviolet Luminous Galaxies (UVLGs), as defined by \citet{heckman05}, with $l_{FUV} \geq 10^{8}$ L$_{\odot}$\ kpc$^{-2}$ that classifies them in the "compact" category. Figure \ref{lfuv_r} illustrates the compacity of our galaxies among the UVLGs and LBGS, with their respective classification criteria on $L_{\rm FUV}$\ and $l_{\rm FUV}$ overplotted. It is instructive to note, in reference to the present discussions related to local objects and the implications on high-z observations discussed later, that Haro 11 and IRAS 08339+6517 , could, regarding also their SFR and metallicity, be considered as UVLGs and LBG analogs.  
\paragraph{}
One of the consequences of the decoupling of Ly$\alpha$\ from non-resonant radiation is the diffuse emission halo observed in all our sample. We present in Fig. \ref{morph_plots} the fraction of this diffuse component  with respect to the total Ly$\alpha$\ flux as a function of the equivalent radius from the brightest emission source which has been isolated in each galaxy to produce the scatter plots of Sect. \ref{analysis_sec}. The radius is yet derived from the surface size covered by the emission. SBS 335-052 and Tololo 65 are not represented on the figure since they do not show any bright (direct) emission that might be identified as photons production sources and are only dominated by a weak diffuse emission. At a distance of 2 kpc, the Ly$\alpha$\ emission reaches  40\% of the total galaxy emission in only one galaxy, where for the remaining ones, it is still around 20\% or below. We need to integrate until 3  to 10 kpc (depending on galaxies) to account for the whole Ly$\alpha$\ flux. These considerations are relative to the threshold level reached by the observations. We consider, because of the masking technique, that above 80\% of the total emission we are background limited. However, we can infer from these plots that the bulk of Ly$\alpha$\ emission consists of a low surface brightness region. 

The situation is slightly different in the galaxies not plotted for which the decrease of the surface brightness with radius is less steeper for reasons stated above. This illustrates the domination of the photon diffusion mechanism in the escape of Ly$\alpha$\ photons, which appears at low surface brightness and extends to large physical scale (several kpc).

\subsection{Age and evolutionary effects}

\begin{figure}[!htbp]
 \centering 
 \includegraphics[width=4cm,height=4cm]{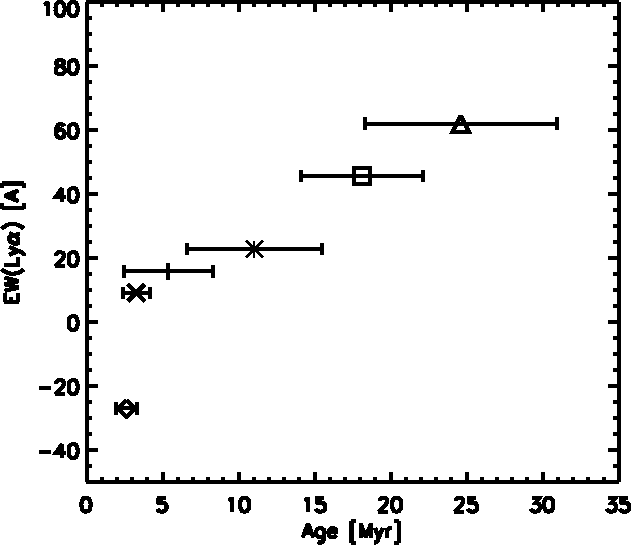}
 \includegraphics[width=4cm,height=4cm]{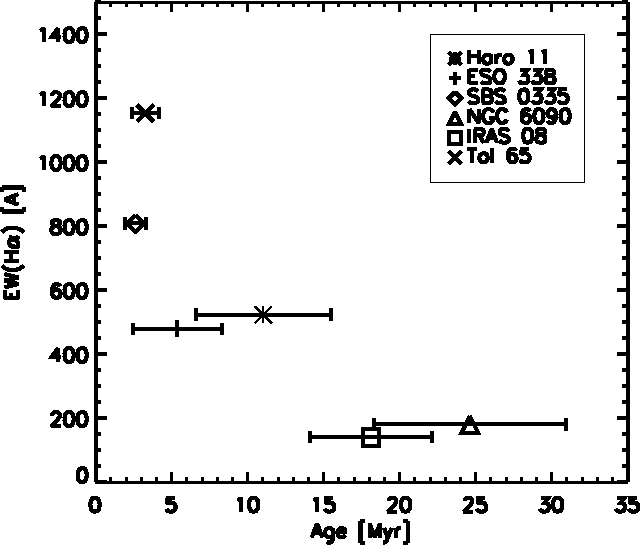}
 \caption{Age correlations: The \text{left} plot shows the integrated $EW_{\mathrm{Ly}\alpha}$\ against the mean age integrated over the galaxy. On the \textit{right}, $EW_{\mathrm{H}\alpha}$\ is plotted as a function of the age. 1-$\sigma$ errors on the age fit are also shown. See text for age calculation and errors estimate. Each point represents a galaxy and the legend is given in the inset.} 
 \label{gen_ages}
 \end{figure}

We have computed the mean age for each galaxy from the SED fitting output in each resolution element with two stellar components as free parameters assuming in both of them an instantaneous burst, and averaging over the whole integration aperture by weighting the age in each resolution element by the corresponding H$\alpha$\ luminosity: 
\begin{equation}
mean~age = \frac{\sum_{i} L_{H\alpha,~i}  \times age_{i}} {\sum_{i} L_{H\alpha,~i}} 
\label{mean_age_eq}
\end{equation}
except for Tololo 65, for which no HST H$\alpha$\ image is available and the luminosity in the B band is used for the weighting. We also carried out Monte Carlo simulations in order to estimate the errors on the weighted age. Each pixel has been resampled with 1000 data points and the fitting procedure is applied to this new sample. The standard deviation gives the 1-$\sigma$ errors plotted in Fig. \ref{gen_ages}. 

  The equivalent widths of strong hydrogen recombination lines are known to be, in principle, good age indicators since they measure the ratio between young ionising over old non-ionising radiations \citep[][and references therein]{leitherer05}. In Fig. \ref{gen_ages} we observe an anticorrelation between H$\alpha$\ equivalent width and the age of the galaxy. This is what is expected by SED models (starburst99 for instance) for both H$\alpha$\ and Ly$\alpha$\ equivalent widths, for ages $\geq$ 1 Myr. This is due to the decrease of the ionizing photons quantity and the increase of the number of stars that contributes to increase the continuum with time. However the dispersion observed is symptomatic of the complexity of this indicator in practise. $EW_{\mathrm{H}\alpha}$\ can be affected, among other effects, by the difference in reddening between nebular and continuum radiations and also the contribution of underlying older stellar population diluting the continuum.

Plotting the Ly$\alpha$\ equivalent width yields a different result. $EW_{\mathrm{Ly}\alpha}$\ does not follow this evolutionary sequence where an increase is observed as function of the age. The present observations show the additional complexity of Ly$\alpha$\ with respect to H$\alpha$\ radiation, since we are comparing in the same plot different objects with different mechanisms at work, such as expanding shell in some ones and static medium in others, and/or probably clumpy ISM, that makes Ly$\alpha$\ quite difficult to interpret in a general way. Large variations in the observed $EW_{\mathrm{Ly}\alpha}$\ are found in LBGs where most of them have likely the same intrinsic $EW_{\mathrm{Ly}\alpha}$. \citet{schaerer08} have shown that for an extinction of E(B-V)$\sim$ 0.3 an intrinsic emission with $EW_{\mathrm{Ly}\alpha} \geq$ 60 $\AA$ is transformed into an absorption, hence a negative observed $EW_{\mathrm{Ly}\alpha}$. Radiation transfer and dust effects can lead to large differences between intrinsic and observed Ly$\alpha$\ equivalent widths. In addition, the apparition and the evolution of superwinds is yet function of the age, and we can expect that EW$_{\mathrm{Ly}\alpha}$\ rises in the presence of ISM kinematical effects. Because of the limited statistics, this could be considered only as a first step toward a more detailed and significant investigation. It is however worth noting that the small age ($\sim$ 2.6 Myr) of SBS 335-052 agrees with our discussion about this galaxy as a young starburst embedded in a static H{\sc i}\ leading to a damped absorption. This is also true for Tololo 65 ($\sim$ 3.3 Myr) where no direct emission is seen and for which GHRS spectrum shows a damped absorption.

\begin{table*}[!htbp]
\centering

\begin{tabular}{l c c c c c c c c c}

\hline \hline \\
Target        &log(L$_{FUV}$)&log(L$_{FIR}$)& SFR(UV)& SFR(Ly$\alpha$)&SFR(FIR)&SFR(H$\alpha$)&(SFR$_{UV}$/SFR$_{Ly\alpha}$)    &(SFR$_{UV}$/SFR$_{Ly\alpha}$)$_{COR}$ \\
              &              &             &        &                &        &      &        &     \\
\hline \\ 
Haro 11       & 10.3         & 11.1        & 5.1    &  1.07          &21.7    &16.33 & 4.77   & 0.53    \\
ESO 338-IG04  &  9.76        &  9.8        & 1.56   &  0.3           & 1.1    & 3.9  & 5.20  & 3.20    \\
SBS 0335-052  &  9.1         &  9.4        & 0.34   &  ...           & 0.4    & 1.1  & ...   & ...      \\
NGC 6090      & 10.03        & 11.4        & 2.9    &  1.1           &43.3    & 20.7 & 2.64  & 0.015     \\
IRAS 08339+6517     & 10.3         & 11.0        & 4.4    &  2.2           &17.3    & 14   & 2     & 0.85    \\
Tololo 65     &  8.3         &  8.4        & 0.05   &  0.008         &0.05    & 0.24 & 6.25  & 2.6     \\
\hline
\end{tabular}
\caption{Star formation rates and luminosities. L$_{FUV}$ is calculated with $\lambda \times f_{\lambda}$ where $\lambda$=1525\AA. $f_{\lambda}$ is the flux density in F140LP filter. L$_{FIR}$ is from \citet{ostlin08}. Both are expressed in units of L$_{\odot,Bol}$ (3.8$\times$10$^{33}$ ergs s$^{-1}$). Star formation rates are derived from the integrated fluxes over apertures based on sky background threshold and using the calibration of Kennicutt (1998) (presented in units of $M_{\odot}~\mathrm{yr}^{-1}$). All quantities are corrected for galactic foreground extinction. The last column ratio only is corrected for internal reddening using $E_{B-V,\mathrm{stars}}$\ for SFR$_{\rm UV}$ correction and $E_{B-V,\mathrm{gas}}$\ for SFR$_{{\rm Ly}\alpha}$ one. }
\label{sfr_table}
\end{table*}

\subsection{Reddening correction and star formation rate} 
\label{sfr_sec}
\begin{figure}[!htbp]
 \centering 
 \includegraphics[width=7cm,height=6cm]{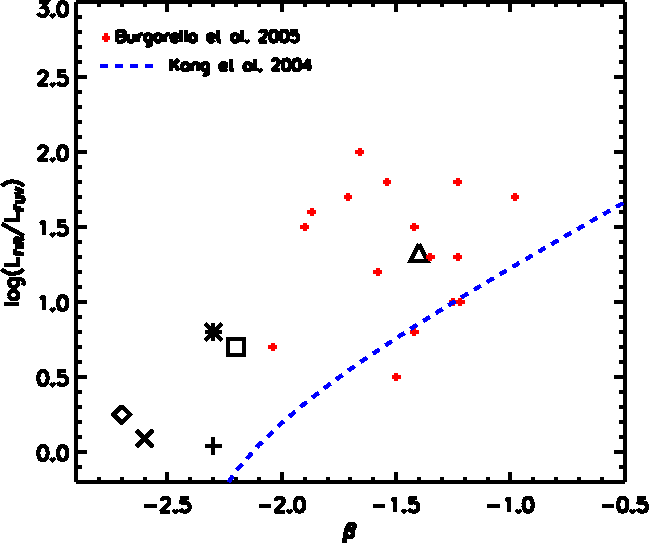}
 \caption{L$_{IR}$ to L$_{FUV}$ ratio versus the UV slope $\beta$. Our galaxies are plotted with the same symbols as in previous figures. $\beta$ is determined from the SED fitting. The remaining red points are galaxies from ELAIS S1 observations in \citet{burgarella05}. The law describing the relationship between FIR to FUV luminosity ratio and the slope $\beta$, deduced from UV observations \citep{kong04} is overplotted (dashed blue line).}
 \label{firfuv}
 \end{figure}

 The evolution of the $L_{\rm FIR}$/$L_{\rm FUV}$\ ratio provides an alternative to evaluate the reddening estimate reliability. We show in Fig. \ref{firfuv} a weak correlation between this luminosity ratio and the UV continuum slope $\beta$ derived from the SED fit. 
We have plotted on the same figure (in blue) the predicted relationship \citep{kong04} between $L_{\rm FIR}$/$L_{\rm FUV}$\ and $\beta$ following \citet{meurer99}. Our galaxies have the same dispersion behavior as those of \citet{burgarella05}, overplotted as red points, which are though more dusty and luminous in the IR. NGC 6090 falls among the red points as it is classified as LIRG and is a very dusty starburst with $E_{B-V,\mathrm{gas}}$\ $\sim$ 0.75). Three of our galaxies (NGC 6090, Haro 11 and IRAS 08339+6517 ) appear to be Luminous Infrared Galaxies (LIRGs, log($L_{\rm FIR}$) $>$ 11 L$_{\odot}$). In contrast with IUE starburst sample of \citet{meurer99}, \citet{goldader02} found, as for our galaxies, that LIRGs and ULIRGs of their sample fall above the line. \citet{seibert05} have also derived, using a large sample of galaxy types, a correction that lowers the empirical $L_{\rm FIR}$/$L_{\rm FUV}$-$\beta$\ reddening relation and is thus also in disagreement with our observations. The observed discrepancy suggests that a simple empirical law, is actually not representative of the observations and the galaxies suffer apparently a higher attenuation than suggested by $\beta$.
Star-formation rate is an essential diagnostic tool of the evolution of galaxies. It allows, since the high-redshift universe has became reachable, to study star-formation episodes and their evolution on a wide range of epochs \citep[e.g.][]{madau96, giavalisco04}. Many indicators that ranges from rest-frame UV to infrared are used to estimate the SFR, and a commonly used one in the distant universe is based on Ly$\alpha$\ emission, since the window opened on the high-z candidates ended up with using the Ly$\alpha$\ line as the main detection and investigation tool. One of the most critical issues related to estimate the SFR, from UV or optical indicators, is the correction for internal reddening. One needs to properly estimate the dust obscuration and therefore the intrinsic flux to make a correct conversion to star-formation rate. This issue is even more critical for the indicator based on Ly$\alpha$\ emission, considering the radiative transfer complexity of this line discussed in this paper. Table \ref{sfr_table} summarizes the SFRs computed from different indicators using \citet{kennicutt98} calibrations. The conversion from flux to SFR from these calibrations assumes a continuous star formation regime in the equilibrium phase, whereas in our SED fitting procedure, used for the Ly$\alpha$\ continuum subtraction or age estimation, we have assumed an instantaneous burst scenario. Nevertheless, translating our results into star-formation rates gives a useful comparison with previous works using these widely used calibrations.

\begin{figure*}[!htbp]
 \centering 
  \subfigure[SFR$_{Ly\alpha}$ \textit{vs} SFR$_{UV}$]{
 \label{sfr_a}
 \includegraphics[width=8.5cm,height=7cm]{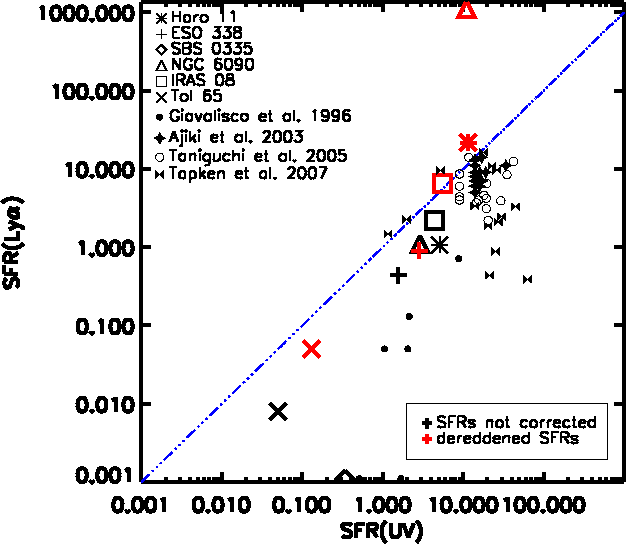}
  \hspace{0.5cm}}
  \subfigure[SFR$_{TOT}$ \textit{vs} different SFRs]{
  \label{sfr_b}
 \includegraphics[width=8.5cm,height=7cm]{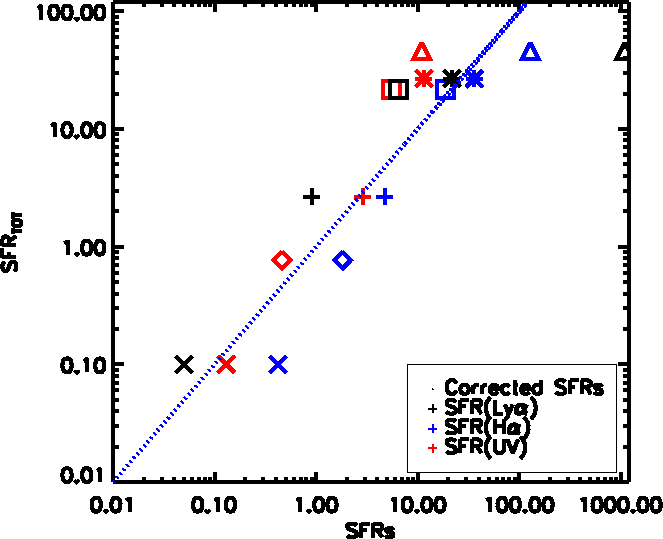}}
 \caption{Star formation rates: Figure \ref{sfr_a} shows SFR derived from nebular emission line Ly$\alpha$\ versus SFR derived from UV continuum. The dark points represent undereddened SFRs. Note that because of the logarithmic scale and for the sake of readability, we have put galaxies for which SFR$_{Ly\alpha}$=0 at SFR$_{Ly\alpha}$=0.001. Observed SFRs from literature at low-z \citep[][and IUE data from NED]{giavalisco96} and high-z \citep[][for LAEs]{taniguchi05, ajiki03} \citep[][for LBGs]{tapken07} are also overplotted and explained in the legend. For our sample, the red points represent SFR$_{Ly\alpha}$\ dereddened using $E_{B-V,\mathrm{gas}}$\  and SFR$_{UV}$ dereddened using $E_{B-V,\mathrm{stars}}$. The dashed line is for SFR$_{Ly\alpha}$\ $=$ SFR$_{UV}$. Figure \ref{sfr_b} represents SFR(UV+IR) (where SFR$_{UV}$\ is not corrected for extinction) versus different corrected SFRs. The legend is also marked on the figure. Units are in M$_{\odot}$~yr$^{-1}$.}
 \label{sfr_tot}
 \end{figure*}

UV emission from galaxies traces the young stellar population and a conversion from UV luminosity to SFR can be computed across the UV range (1250-2500 \AA). This estimation is very sensitive to dust attenuation, because of the wavelength domain and the patchy ISM \citep{kennicutt98}\footnote{SFR$_{\mathrm{UV}}$(M$_{\odot}$~yr$^{-1}$) = 1.4 $\times$ 10$^{-28}$~L$_{\nu}$(ergs~s$^{-1}$~Hz$^{-1}$)}.
 Alternatively, nebular emission lines, such as Ly$\alpha$, provides information on the ionising flux of the young massive stellar population with lifetimes $<$ 20 Myr. Therefore, it provides a quasi-instantaneous estimation of the current star-formation rate. For case B recombination theory \citep{brocklehurst71} we have: 
\begin{equation}
\label{sfrlya_eq}
\mathrm{SFR}_{\mathrm{Ly}\alpha}(M_{\odot}~\mathrm{yr}^{-1}) = 9.1 \times 10^{-43}~L(\mathrm{Ly}\alpha)~(\mathrm{ergs~s}^{-1})
\end{equation}

Yet, in addition to the IMF dependency, this method is highly sensitive to the extinction correction as we have mentioned before. Figure \ref{sfr_tot} is illustrative of the reddening correction issue. Figure \ref{sfr_a} represents SFR(Ly$\alpha$) versus SFR(UV): for dark points neither SFR is corrected and for red points SFR(Ly$\alpha$) is corrected using $E_{B-V,\mathrm{gas}}$\ and SFR(UV) using $E_{B-V,\mathrm{stars}}$\ . In the purpose of comparison, data from literature are also included. It is interesting to note the consistent discrepancy between SFR(Ly$\alpha$) and SFR(UV) (dark points) for both our sample and objects from literature, including low- \citep{giavalisco96} and high-redshift galaxies \citep{taniguchi05, ajiki03, tapken07}. The lack of points under the line of equality at low SFR(Ly$\alpha$) for high-z observations, is a consequence of the completeness limit, for LAEs in particular, excluding faint Ly$\alpha$\ emitters. Indeed, we expect, because of the resonance effects of Ly$\alpha$ to observe a scattered distribution below the line of slope unity providing one uses a statistically significant sample. Such a distribution would help to characterize a potential upper limit of the discrepancy between resonant and non-resonant SFR indicators. Such a diagram could also serve as an important probe for the galaxy evolution from the damped and/or young systems lying at SFR$_{{\rm Ly}\alpha} \sim 0$, to more evolved starburst events with higher ionised gas fraction and/or undergoing feedback outflows hence approaching the line where SFR$_{{\rm Ly}\alpha} \sim$~SFR$_{\rm UV}$.  

For our galaxies, in the case of observed values, SFR(Ly$\alpha$) is systematically below the equal values line, underestimating SFR by a factor between 2 to 6 with respect to SFR(UV). This discrepancy, usually observed in high-z galaxies also, emphasizes the highest attenuation of Ly$\alpha$\ emission line with respect to UV continuum. The correction for dust attenuation, one of the main responsible of this discrepancy, brings the two star-formation rates in a better agreement except for NGC 6090. 


  When SFR derived from the UV is not corrected for absorption, the SFR measured from the infrared should, in principle, be complementary, since radiation that is strongly absorbed in the UV is re-emitted in the thermal IR \footnote{ SFR$_{\mathrm{FIR}}$(M$_{\odot}$~yr$^{-1}$) = 4.5 $\times$ 10$^{-44}$~L$_{FIR}$(ergs~s$^{-1}$) }.
Figure \ref{sfr_b} shows the "total" SFR (SFR$_{\rm UV}$\ + SFR$_{\rm FIR}$) where SFR$_{UV}$\ is not corrected, versus dereddened SFR$_{Ly\alpha}$\ and SFR$_{H\alpha}$ with $E_{B-V,\mathrm{gas}}$ (dark and blue points),  and SFR$_{UV}$\ with $E_{B-V,\mathrm{stars}}$ (red points). The corrected SFR(Ly$\alpha$) is still below the total SFR for most of the galaxies except for NGC 6090 where the dust correction highly overestimates the total SFR (we have discussed possible reasons for this overestimation in Section \ref{role_dust_sec}). Conversely, SFR derived from corrected H$\alpha$\ (blue points) luminosity yields a better result. Similarly, the dereddened UV estimator (red points) places the galaxies rather close to the line of equality.

Finally the different results found when deriving  SFR from Ly$\alpha$\ and others indicators such as H$\alpha$\ or UV, and in particular the failure of Ly$\alpha$\ indicator to recover the total SFR (UV + IR) even when corrected for reddening, are indicative of the decoupling of resonant Ly$\alpha$\ and non resonant (e.g. UV continuum or Balmer lines) radiation with respect to the dust obscuration and therefore the difficulty to use this line as a reliable star formation indicator.

As an alternative, we can take advantage of the available informations of our observations to make a better estimate of the star formation rate when only Ly$\alpha$ luminosity is known. We can calculate the Ly$\alpha$\ escape fraction using the corrected H$\alpha$\ flux and assuming the case B recombination ratio Ly$\alpha$/H$\alpha$:
\begin{eqnarray}
\label{f_esc_eq}
f_{esc}(Ly\alpha) &=&  f(ly\alpha) / (8.7 \times f(H\alpha)_C) \\
f(H\alpha)_C~~ &=&  f(H\alpha) \times 10^{(1.048\times E(B-V)_{gas})}  
\end{eqnarray}
 where f(ly$\alpha$) is the observed flux and $f(\mathrm{H}\alpha)_C$\ is the H$\alpha$\ flux corrected pixel by pixel for internal reddening using \citet{cardelli89} extinction law. Unlike high-z observations, where in the best case only the global UV slope is known, our present study gives access to accurate extinction information. Therefore, f$_{esc}$ is a good estimate of the intrinsic Ly$\alpha$\ flux, since it takes into account both dust obscuration and resonant scattering mechanism and hence allows us to correct the SFR$_{Ly\alpha}$\ calibration (Eq. \ref{sfrlya_eq}) as follows:
\begin{equation}
\label{sfrlyacor_eq}
\mathrm{SFR}_{\mathrm{Ly}\alpha}(M_{\odot}~\mathrm{yr}^{-1}) = (1/f_{esc}) \times 9.1 \times 10^{-43}~L(\mathrm{Ly}\alpha)~(\mathrm{ergs~s}^{-1})
\end{equation}
The escape fractions obtained for the six galaxies are reported in Table \ref{ebv}. We can assume, in general terms, an escape fraction of 5\% as a mean "statistical" value to deduce the appropriate SFR when f$_{esc}$ is not available. This correction is evidently subject to uncertainties due to f$_{esc}$ variations, but is in any case more representative of the reality than standard calibrations \citep[][for instance]{kennicutt98}.

The revised star formation rate based on $L(\mathrm{Ly}\alpha)$ is equivalent, according to the definition of the escape fraction in Eq. \ref{f_esc_eq}, to SFR(H$\alpha$) corrected for reddening. Hence, it is represented on the right plot of Fig. \ref{sfr_tot} by the blue points, which appears to give a better estimation of the total SFR than that given by the SFR$_{Ly\alpha}$\ corrected for only dust obscuration.

\subsection{Implications for high-redshift galaxies}

\paragraph{}

\begin{figure}[!htbp]
 \centering 
 \includegraphics[width=7cm,height=6cm]{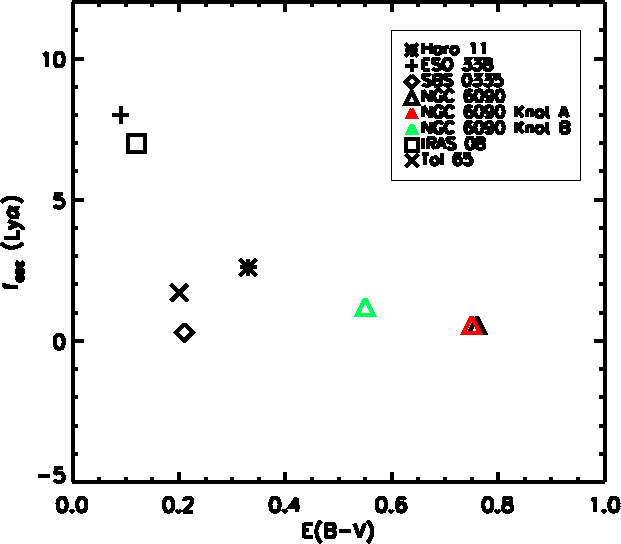}
 \caption{Escape fraction of Ly$\alpha$\ photons (in percent, see text for details on the f$_{esc}$ determination) as a function of extinction E(B-V) in the gas phase. We have included the two components (knots A and B) of the interacting system NGC 6090 and treated them separately by a masking procedure. We observe a decline in the amount of escaping photons when increasing the dust amount. We note also that most of the galaxies have a small f$_{esc}$\ around 3\% or below.} 
 \label{fesc}
 \end{figure}

We show in figure \ref{fesc} how $f_{esc}$ is function of the extinction. We observe that the Ly$\alpha$\ escape fraction is somewhat correlated with $E_{B-V,\mathrm{gas}}$\ but with an important dispersion in, moreover, a quite limited sample. At a galaxy scale we see the manifestation of the dust obscuration effect on emergent Ly$\alpha$\ radiation, although in a purely dust regulated model, this correlation would have been more striking. It is also interesting to note that correcting the total H$\alpha$\ flux using the mean $E_{B-V,\mathrm{gas}}$\ lead to higher escape fractions by 15 to 40 \% that those calculated using pixel level corrections. In high redshift observations only global correction is possible in absence of spatial resolution. 

The rate of escaping Ly$\alpha$\ photons does not exceed 10\% in this sample. In addition, we have outlined the ubiquitous halo of diffuse emission present in all the observed galaxies, and we are able to get a feeling of how important is this contribution in section \ref{morph_sec}. The diffuse component represents the bulk of the Ly$\alpha$\ emission (Table \ref{ebv}). It is very likely that such low surface brightness emission remains undetectable in high-redshift galaxies. This point out the caution that should be taken when deriving physical quantities, such as star formation rates (SFRs), from Ly$\alpha$\ alone, since we are dealing with only a small fraction of escaping photons, and we are probably missing the majority of this fraction. Simple extinction correction of SFR could also be at fault since such method has failed to recover the value determined by the recombination theory in most cases and the behavior of resonant radiation according to the dust content is unpredictable without any complementary information (HI distribution, gas kinematics, etc.). However a better, and more realistic estimate could be obtained using the calibration proposed in Eq. \ref{sfrlyacor_eq}.

Because of the difficulties discussed above, high-z star formation rates  based on Ly$\alpha$\ are generally underestimated with respect to that derived from UV for instance. Discrepancies have been observed between the two estimation methods where SFR based on the Ly$\alpha$\ luminosity was smaller by a factor of two or more than that based on the UV continuum \citep{hu02, kodaira03}. \citet{taniguchi05} found that SFRs derived from Ly$\alpha$\ for their sample of z $\sim$ 6.6 LAEs lie a factor of 5, on average, below those based on UV continuum. \citet{tapken07} found similar discrepancy for their UV-selected galaxies. This is also what is seen at low redshift where a factor of 2 to 6 is found (the present work, Section \ref{sfr_sec}). 

Ly$\alpha$\ emission line has became a powerful tracer of star formation at high-redshift. However, it is clear that using only the Ly$\alpha$\ luminosity or equivalent width for the characterization of star formation episodes leads to many uncertainties which inhere in the Ly$\alpha$\ radiative transfer complexity and the uncertainties of the SFR calibration methods.
Despite differences due to obvious evolution effects between local and high-z star-forming galaxies \citep[see][]{mashesse03}, the physical processes governing the radiative transfer and the escape mechanism of Ly$\alpha$\ photons should be the same, and motivate the present extrapolation. Nevertheless, we stress again the limited number of galaxies in our sample that precludes any statistically significant study and prompt a further investigation with an extended sample of galaxies and complementary spectroscopic study.

\begin{table}[!htbp] 
\centering
\begin{tabular}{l c c c c c c}

\hline \hline \\
Target           &  &f$_{esc}$ &f$_{diffuse}$ & E(B-V)         & E(B-V)            & AGE    \\  
                 &  &(\%)      & (\%)         & \textit{gas}   & \textit{stars}     & (Myr)  \\
\hline \\
Haro 11          &   & 2.6     &  74          & 0.33           &  0.07              &  11    \\
ESO 338-IG04     &   & 8       &  70          & 0.08           &  0.08              &  5.4    \\
SBS 0335-052     &   & 0.25     & $\sim$ 100   & 0.21           &  0.04              &  2.6    \\
NGC 6090         &   & 0.56    &  73          & 0.76           &  0.18              &  25     \\
IRAS 08339+6517        &   & 7       &  65          & 0.12           &  0.03              &  18     \\
Tololo 65        &   &1.7      & $\sim$ 100    & 0.2            &  0.12              &  3.3     \\
\hline
\end{tabular}
\caption{Integrated properties and Ly$\alpha$\ emission characteristics. quantities are integrated over the same apertures defined in table \ref{catalog}. The first column contains the escape fraction of Ly$\alpha$\ photons for each galaxy assuming a case B Ly$\alpha$/H$\alpha$\ recombination ratio (see text for details). For SBS 0335-052, only positive Ly$\alpha$ contribution is used to derive an upper limit for f$_{esc}$. Second column represents the contribution of the diffuse component in the whole Ly$\alpha$\ emission determined by masking the bright "direct" emission sources if any. The mean nebular extinction $E_{B-V,\mathrm{gas}}$\ is derived from the ratio of integrated H$\alpha$\ to H$\beta$\ fluxes and is showed in column 3. The fourth gives the continuum extinction determined from the SED fitting procedure. Last column shows the age issued from the SED fitting and averaged on the total galaxy pixels following Eq. \ref{mean_age_eq}.} 
\label{ebv} 
\end{table}

\section{Conclusions}
\label{conclusion_sec}
\paragraph{}
 Combining space (HST) and ground-based (NOT and NTT) observations we have mapped the Ly$\alpha$\ emission and the dust content in six nearby star-forming galaxies. We have compared the extinction E(B-V) produced from the Balmer decrement H$\alpha$/H$\beta$\ to several parameters like Ly$\alpha$\ emission, equivalent width or recombination ratio Ly$\alpha$/H$\alpha$\ at a small scale in order to disentangle the role of the dust from other parameters. Implications for high-z studies, via the global properties of the galaxies have also been investigated.

Our galaxies exhibit different Ly$\alpha$\ morphologies from emission to damped absorption or combination thereof:
 
\begin{itemize}
\item{ In systems with emission and absorption (namely Haro 11 and ESO 338-04) we found Ly$\alpha$\ photons emerging from regions with similar or even higher extinction than those where Ly$\alpha$\ is seen in absorption. We point out the role of the ISM distribution, where in the case of clumpyness morphology Ly$\alpha$\ photons escape preferentially to H$\alpha$\ ones leading to an observed Ly$\alpha$/H$\alpha$\ ratio higher than the theoretical level corrected for extinction. 
 
}

\vspace{0.2cm}
\item{ In objects that show no strong absorption (NGC 6090 and IRAS 08339+6517 ) we observe no clear correlation between Ly$\alpha$\ and the dust content. The H{\sc i}\ kinematics may play a more significant role in the escape of Ly$\alpha$\ photons as confirmed by kinematics studies which have shown large ISM outflows in both systems.}

\vspace{0.2cm}
\item{ SBS 335-052 is a Ly$\alpha$\ absorber with a great H{\sc i}\ column density coverage which is believed to be static with respect to the emitting region. We estimate an age ($<$ 5 Myr) in agreement with the picture where the starburst is too young to have ionised the surrounding gas or driven an outflow. We observe precisely what is expected from the resonant nature of Ly$\alpha$\ in a static neutral gas: a damped absorption with a declining relationship between Ly$\alpha$\ and E(B-V) indicating that the dust is, in this case, the main regulator of Ly$\alpha$\ escape.}

\end{itemize}
 
 When investigating global parameters of our sample, we found that simple dust extinction correction fails to recover the intrinsic Ly$\alpha$/H$\alpha$\ ratio, where the role of the dust is, in some cases, underestimated because of the resonant scattering, and in other cases, overestimated because of the clumpyness distribution of the ISM. We observe neither no evident correlation between $EW_{\mathrm{Ly}\alpha}$\ and the reddening. The observed Ly$\alpha$\ escape fraction is found to not exceed 10\% in our sample and is, for most of our galaxies, about 3\% or less.

The resonant decoupling of Ly$\alpha$\ from non-resonant radiation leads also to an ubiquitous diffuse halo with low surface brightness. It represents the bulk of the Ly$\alpha$\ emission and extend to regions at several kpc from emitting regions and which are not reached by H$\alpha$\ or continuum radiation yielding high $EW_{\mathrm{Ly}\alpha}$.

Because of the radiative transfer complexity of the Ly$\alpha$\ line, star formation rate (SFR) based on Ly$\alpha$\ lead to different results than SFR derived from other indicators (from UV for instance), and fails to recover the total SFR (UV + IR), and this even when corrected for dust obscuration, preventing any determination of the intrinsic star formation rate. We propose therefore a more realistic calibration for the SFR when information on Ly$\alpha$\ only is available (which is usually the case for high-redshift surveys), accounting for dust attenuation and resonant scattering phenomenon through the Ly$\alpha$\ escape fraction.


\begin{acknowledgements}
We are very grateful to Daniel Schaerer for thoughtful and valuable comments on the draft of this paper. We thank Anne Verhamme for useful discussions, Claus Leitherer and Artashes Petrosian for their work on the Ly$\alpha$\ project.  
Filter 113 for ALFOSC observations was acquired thanks to a grant from Erik Holmberg foundation.
\end{acknowledgements}

\bibliographystyle{aa}
\bibliography{references}

\end{document}